# Modular Design and Experimental Evaluation of 5G Mobile Cell Architectures Based on Overlay and Integrated Models

José Ruela, Ivan Cojocaru, André Coelho, Rui Campos, Manuel Ricardo

INESC TEC, Faculdade de Engenharia, Universidade do Porto, Portugal
jose.ruela@inesctec.pt, up201907151@edu.fe.up.pt, andre.f.coelho@inesctec.pt,
rui.l.campos@inesctec.pt, manuel.ricardo@inesctec.pt

**Abstract.** This paper presents the concept, architectural design, and performance evaluation of a 5G Mobile Cell (MC) used to provide 5G wireless connectivity to User Equipment (UE) in areas with limited fixed 5G infrastructures or subject to adverse radio conditions. We consider two main approaches to MC design: an overlay model, where the MC obtains backhaul connectivity from a 5G overlay network, and an Integrated Access and Backhaul (IAB)-based model, discussing their protocol stacks and architectural implications. In order to validate the MC's performance, we employ an emulation-based testbed using the OpenAirInterface (OAI) implementation, considering different MC positions. The results validate the MC concept and demonstrate that MC positioning significantly influences network performance. This paper has the potential to aid network operators and service providers in selecting and deploying MC architectures for temporary coverage extension and capacity reinforcement in different environments, including seaports, industrial scenarios, and public safety.

## 1 Introduction

### 1.1 Mobile Cell – concept and goals

A 5G *Mobile Cell* (MC), as proposed and described in this paper, is a communications system installed in a mobile platform that provides 5G mobile terminals with access to subscribed 5G network services, using the same standard procedures as if they were connected through a stationary 5G *Base Station*, which is designated *next generation NodeB*, abbreviated to gNB.

The first goal of a 5G MC was to extend or improve the coverage of a public or private 5G network [1]. The MC provides 5G connectivity to mobile terminals (inside or in the vicinity of the moving vehicle) when they are not within radio reach of the 5G access network or when adverse propagation conditions severely degrade the quality of communication or even prevent it (e.g. due to permanent or temporary obstacles).

In a broader context, an MC may be a component of emerging *on-demand wireless networks* [2]. These networks are characterized by dynamic reconfiguration and repositioning capabilities and offer a suitable solution to extend, restore or reinforce wireless coverage where a permanent access infrastructure does not exist, is insufficient or damaged, such as in temporary events [3] or public safety emergency scenarios [4]. In some cases, an MC may be deployed as a nomadic node placed in a predefined (quasi) static position where it stays while needed.

### 1.2 Context

The design and development of a 5G MC prototype, as well as its deployment and evaluation in a real 5G network environment as a proof of concept is being carried out in the context of the NEXUS [5] project funded by the *NextGenerationEU*, within the scope of the Recovery and Resilience Mechanism (MRR) of the European Union (EU).

The NEXUS *5G Connected Port* work package aims at providing a private 5G-based network infrastructure capable of supporting a services platform that meets heterogeneous and even hard real-time requirements of present and future key use cases in a seaport environment.

Novel technological solutions, such as mission-critical communications, video surveillance, and autonomous vehicles, are already being used in seaports, fostering environment-aware, data-driven aided decisions as well as real-time, adaptive management of seaport operations [6], thus improving efficiency and reducing costs.

In this context, legacy wireless and wired communications solutions have some limitations. For example, Wi-Fi is prone to interference from other systems operating in the same frequency bands, such as radars of ships while docking; this leads to communications disruptions, which reduce network reliability and service availability, and degrade performance. On the other hand, wired-based technologies offer limited flexibility in dynamic networking scenarios.

The current trend to support advanced services and applications in seaports [6] is to deploy a 5G network infrastructure [7] [8], not only to overcome the limitations of legacy solutions, but also due to advanced features enabled by novel architectural concepts and principles.

First, unlike Wi-Fi, 5G networks operate in a licensed, interference-free spectrum.

Secondly, the 5G *service-based architecture* allows virtualization of network functions, which provide services to each other, and use of software-defined network techniques.

Thirdly, the *network slicing* concept allows flexible creation and customized configuration of independent logical networks with associated physical and logical resources, over the same physical infrastructure. Three standard slice types were defined to represent three generic classes of use cases with different *Quality of Service* (QoS) requirements, which are characterized by configurable values of key performance indicators, such as user data rate, latency, reliability and service availability:

- *enhanced Mobile Broadband* (eMBB) is intended for applications that require high data rates, such as video surveillance, 3D high-definition video or virtual reality.
- *Ultra-Reliable Low-Latency Communications* (URLLC) is targeted at applications that have stringent delay requirements and need very high reliability and availability, such as video-based remote control (e.g. cranes), mission-critical services, or autonomous vehicles.
- *Massive Machine-Type Communications* (mMTC) is optimised to handle a large number of devices that generate low volumes of non-time-critical data, such as smart metering or the collection of sensor data.

The use of an MC in seaports allows its functional and performance evaluation in different radio propagation conditions, since it can be carried, for example, by ground vehicles, towboats, or drones (tethered or not), as illustrated in Figure 1 [9].

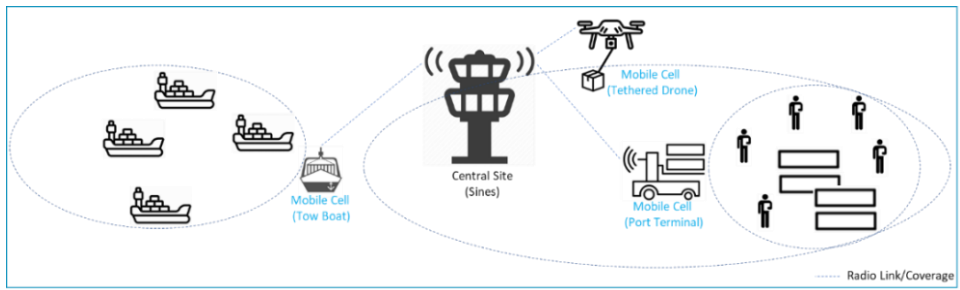

Figure 1. Mobile Cell carried by ground vehicles, towboats, or drones.

In general, an MC may be deployed in a broad scope of application scenarios characterized by a diversity of use cases that may be mapped to 5G standard slice types, by adequately configuring the values of the respective key performance indicators.

## 1.3    Design methodology

The MC is a mobile node in a 5G *Radio Access Network* (RAN), which is composed of stationary gNBs.

Both MCs and gNBs exchange control and data messages with *User Equipment* (UE) and *5G Core Network Functions* to assist and manage the communication of mobile users.

An MC must implement a set of functions (strictly speaking, those hosted by the mobile system) and support interfaces with UE and with RAN and Core functional entities (physical or logical).

As a pre-requisite, the MC interfaces and respective protocol stacks must comply with *3rd Generation Partnership Project* (3GPP) 5G Technical Specifications. Moreover, communication between the MC and fixed RAN and Core entities must be supported over a 5G radio access link, using procedures specified by 3GPP.

Under these conditions, the design of the MC depends on two aspects:

- The decision concerning the logical split and physical distribution of gNB functions.
- The choice of the architectural model that provides IP connectivity between the MC and the fixed infrastructure.

First, an analysis and an evaluation of possible MC solutions that combine both aspects (functional split and IP connectivity model) were carried out.

At that stage, 3GPP had already specified the *Integrated Access and Backhaul* (IAB) architecture. Developing an MC based on this model appeared to be a promising choice, especially because IP connectivity is natively supported by IAB networks.

Although an IAB-based *Mobile Base Station Relay* (MBSR) was already being studied and later specified by 3GPP, this solution was not selected for the short-term, due to the lack of an IAB network to support MC development and testing.

As an alternative to the *integrated model*, we decided that, in the short-term, MC development should be based on the *overlay model*. This means that IP connectivity is established by and across a 5G *overlay network* (RAN and Core), which is logically distinct, in its role, from the MC's home network. Properties of both models will be discussed in detail.

Two gNB functional splitting options were considered for development: the MC may host all gNB functions or only a subset, in which case it acts as a *mobile relay*. At a later stage, this MC relay based on the overlay model may be enhanced and evolve to an IAB MBSR.

Recently (March 2024), 3GPP approved a new study item aimed at specifying the *Wireless Access Backhaul* (WAB) architecture, to support mobile gNBs based on the overlay model; conceptually, WAB is similar to the first MC alternative (full gNB).

In this paper, the three candidate MC solutions are fully characterized in architectural terms and compared using a set of criteria, which allow identifying their strong and weak properties. This analysis may be useful when selecting the most adequate solution in different usage scenarios, with specific requirements and constraints.

All references to architectural aspects and related topics will be given in the following sections, in the appropriate context determined by the organization of the paper, which is outlined next.

Section 2 provides background (contextual) information on 5G networks, focused on RAN and gNB architectures, including IAB and IAB-based MBSRs. The design methodology and high-level functional description of the MC solutions based on the overlay model are presented in Section 3. The complete protocol stacks (from UEs to the respective Core Network functions) for the MC solutions under study are derived and discussed in Section 4. Then follows, in Section 5, a comprehensive comparative analysis of these solutions. Section 6 describes the validation of the MC prototypes in a realistic emulation environment. Section 7 summarizes the main contributions and presents conclusions and directions for future work.

## 2    5G networks – background information and related work

### 2.1    5G System architecture – basic concepts

3GPP defined and specified, in a large set of Technical Specifications, a *5G System* (5GS) consisting of a *5G Access Network* (AN), a *5G Core Network* (5GCN), and *User Equipment* (UE).

The 5G AN comprises a *new generation radio access network* (NG-RAN) and/or *non-3GPP ANs* (wired or wireless), which are connected to the 5GCN by different interfaces.

The 5GCN consists of a set of *Network Functions* (NFs), organized in two planes: *Control Plane* (CP) and *User Plane* (UP), as represented in Figure 2.

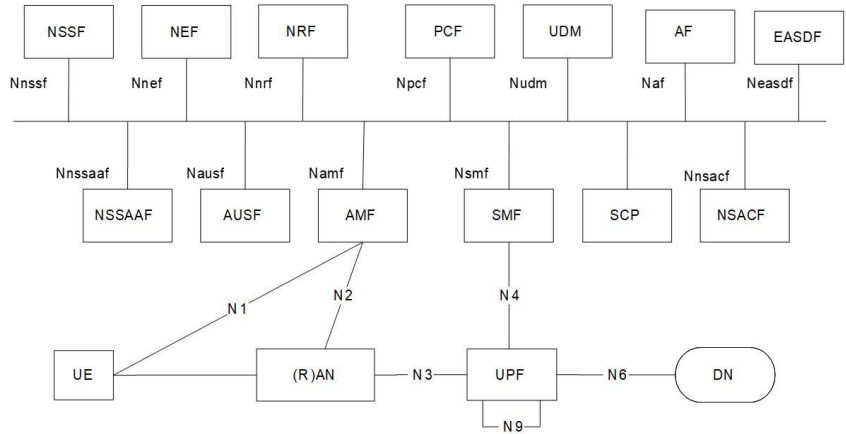

Figure 2. Non-Roaming 5G System Architecture [10].

Interactions between NFs are defined as services (*service based architecture*). The most relevant NFs in this paper are: *Access and Mobility Management Function* (AMF) in the Control Plane and *User Plane Function* (UPF) in the User Plane. A UPF at the edge of the Core Network provides access to *Data Networks* (DNs). Other key CP funtions are *Session Management Function* (SMF), *Policy Control Function* (PCF), *Authentication Server Function* (AUSF) and associated *Unified Data Management* (UDM).

### 2.2    NG-RAN architecture

The NG-RAN uses 5G *new radio* (NR), *Evolved Universal Terrestrial Radio Access* (E-UTRA) / *Long Term Evolution* (LTE) or both as access technologies.

The building block for the NG-RAN logical architecture is the NG-RAN node, which consists of a set of logical functional entities and terminates logical interfaces towards other logical nodes.

As defined by 3GPP in [11], an NG-RAN node can be either a *next generation NodeB* (gNB), i.e. a 5G Base Station, or a *next generation eNodeB* (ng-eNB), which is an enhanced version of an *evolved NodeB* (eNB), i.e. a 4G Base Station. The NG-RAN logical architecture is represented in Figure 3.

NG-RAN nodes are connected to the 5G Core Network (5GCN) via the NG interface and to one another via the Xn interface. Hereafter, we only consider gNB nodes, assuming the *standalone* (SA) mode, that is, the NG-RAN uses 5G NR as access technology and connects to a full 5G Core.

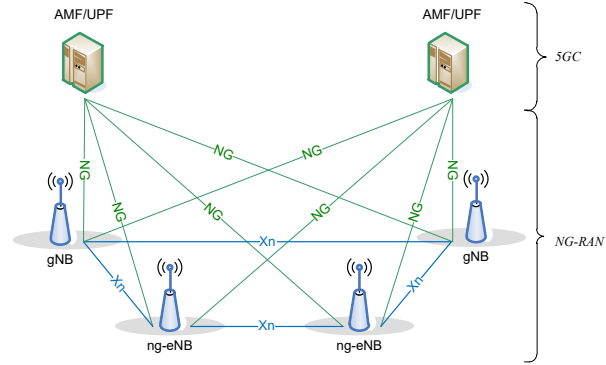

Figure 3. NG-RAN logical architecture – nodes and interfaces [11].

A gNB communicates with *User Equipment* (UE) through the NR Uu interface defined in [11], using the following standard protocols represented in Figure 4: *Radio Resource Control* (RRC), on CP, *Service Data Adaptation Protocol* (SDAP), on UP, and, on both planes, *Packet Data Convergence Protocol* (PDCP), *Radio Link Control* (RLC), *Medium Access Control* (MAC) and *Physical* (PHY).

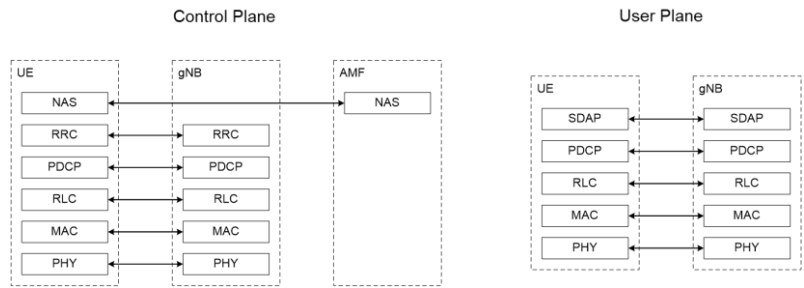

Figure 4. UE and gNB protocol stacks on the NR Uu interface [11].

The *Non-Access-Stratum* (NAS) layer, on CP, between UE and AMF (on the 5G Core) is also represented.

### gNB functional split

Traditionally, RAN architectures were designed based on monolithic building blocks; for example, Base Station functions were implemented by proprietary boxes called *Baseband Units* (BBUs), which were connected to *Radio Units* (RUs), with RF equipment mounted in radio towers.

With the introduction of 5G NR, there was a trend to further disaggregate BBU functions into *Central Units* (CUs) and *Distributed Units* (DUs).

Grouping and organizing gNB layered functions into such logical units (CU, DU, and RU) allows greater flexibility when designing RAN configurations tailored to specific functional and/or performance requirements. Different criteria may be used for this purpose.

On the one hand, it is possible to split logical units on different protocol boundaries. For each boundary (or split), a transport network is necessary to connect the respective units, and an interface must be defined and, possibly, standardized. Latency and throughput requirements on the transport network become more stringent as a split is performed at lower protocol layers.

On the other hand, these units may be deployed on different logical or physical systems, possibly at different locations, in order to achieve more efficient resource sharing among them.

In practice, two split possibilities have deserved attention for standardization purposes: a *High Layer Split* (HLS) and a *Low Layer Split* (LLS). Single or double split architectures are possible.

To evaluate different splitting alternatives, 3GPP conducted a study [12] that analysed eight split possibilities, which are illustrated in Figure 5.

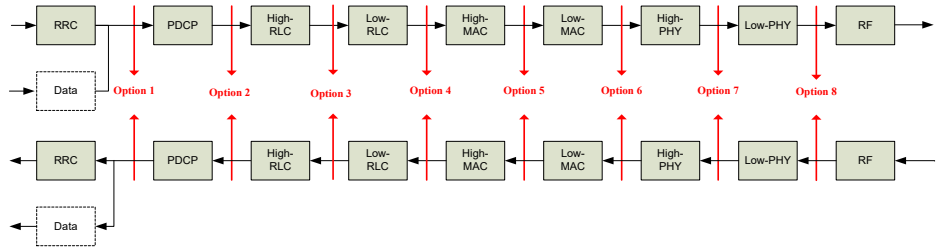

Figure 5. Function Split between central and distributed unit [12].

For its reference NG-RAN architecture, 3GPP selected option 2 for *High Layer Split*, nominally between CU and DU entities, at the boundary between the PDCP and RLC layers, and specified the respective interface (to be detailed). This architecture neither enforces nor excludes a specific *Low Layer Split*.

The *Open RAN (O-RAN) Alliance* specified the double split O-RAN architecture [13] [14]. For HLS, split 2 was adopted as well as the respective 3GPP compliant interface for the communications between an O-CU and one or multiple O-DUs. For LLS, the intra-PHY layer split 7-2x (a variant of split 7) was selected together with the *evolved Common Public Radio Interface* (eCPRI) specification [15] for the *fronthaul* interface between an O-DU and O-RU(s).

For the design of the Mobile Cell, latency and throughput requirements on HLS and LLS interfaces were taken into account.

### gNB architecture – logical nodes

The NG-RAN architecture shown in Figure 6 is defined in [16]. It consists of a set of gNBs connected to the 5GC through the NG interface and interconnected through the Xn interface.

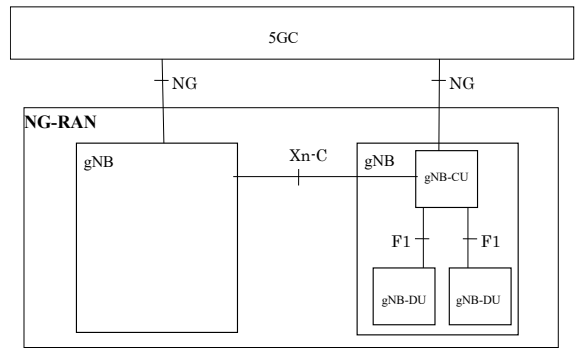

Figure 6. Overall NG-RAN Architecture [16].

The gNB architecture based on split 2 is specified in [16] and consists of a single *gNB Central Unit* (gNB-CU) logical node and one or multiple *gNB Distributed Units* (gNB-DUs) logical nodes.

The gNB-CU and each gNB-DU are connected via the F1 interface and, for control purposes, communicate by means of the *F1 Application Protocol* (F1AP), as will be discussed.

The gNB-CU hosts RRC, SDAP and PDCP protocols and controls the operation of the gNB-DU(s) it is connected to.

A gNB-DU hosts RLC, MAC and PHY protocols and is partly controlled by the only gNB-CU to which it is connected.

The internal structure of a CU/DU split gNB is invisible to UEs, which use the same procedures and execute the supporting protocols as if they were attached to a non-split gNB.

Depending on the logical organization of the lower protocol layers, one gNB-DU may support one or multiple RUs (cells), but one cell is supported by only one gNB-DU.

NG, Xn and F1 are logical interfaces. The NG and Xn-C interfaces terminate in the gNB-CU.

### F1 interface and F1 Application Protocol (F1AP)

The F1 interface was specified to support the communications between the gNB-CU and the gNB-DU(s) that constitute a gNB based on split 2. A transport network is required for this purpose

This transport mechanism is provided by the F1 interface [17]; in particular, the specification defines the F1 interface protocol structure in the Control Plane (F1-C) and in the User Plane (F1-U), and the respective procedures

F1 is an open, logical point-to-point interface between two endpoints (gNB-CU and gNB-DU); thus, a dedicated instance of the F1 interface must be set up and maintained for each (gNB-CU, gNB-DU) pair, even in the absence of a physical direct connection between the endpoints. The F1 interface supports the exchange of signalling information and transmission of user data between the respective endpoints, with CP and UP separation.

As described in [17], the F1 interface separates the *Radio Network Layer* (RNL) from the *Transport Network Layer* (TNL); the F1-C and F1-U protocol stacks are shown in Figure 7. Similar stacks were defined for the NG and Xn interfaces, with NGAP and XnAP as Application protocols, respectively.

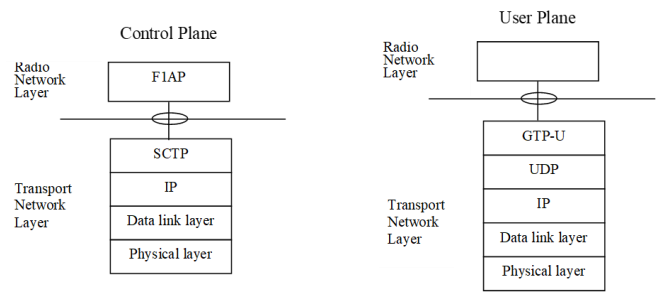

Figure 7. Control Plane and User Plane protocol structure of the F1 interface [17].

On the Control Plane (F1-C), the F1 Application *Protocol* (F1AP) [18] provides the RNL *signalling service* that supports the F1-C functions defined in [17]. F1AP consists of *Elementary Procedures* (EPs), which are units of interaction between a gNB-CU and a gNB-DU. EPs are defined separately and used to build up complete sequences in a flexible manner.

F1AP runs over TNL protocols: *Stream Control Transmission Protocol* (SCTP) [19], IP, and Data Link and Physical layers, which are specific of each transport network.

SCTP provides reliable transfer of F1AP messages over the F1-C interface, networking and routing functions and redundancy in the signalling network, and support for flow and congestion control.

On the User Plane (F1-U), high-level *Protocol Data Units* (PDUs) are transported over TNL protocols: *GPRS Tunnelling Protocol User* (GTP-U) [20], UDP, IP, and Data Link and Physical layers common to F1-C.

F1-U functions support the transfer of user data between a gNB-CU and an associated gNB-DU and the control of downlink user data flow to a gNB-DU.

RRC, SDAP and PDCP protocols terminate on UEs and the gNB-CU. RRC and PDCP PDUs on CP are carried by F1AP, while SDAP and PDCP PDUs on UP are carried by GTP-U.

## 2.3    Integrated Access and Backhaul (IAB) – concept and architecture

*Integrated Access and Backhaul* (IAB) is an extension of the NG-RAN architecture, which provides NR wireless relaying of UE access traffic over backhaul links. The IAB architecture is supported by two types of logical nodes (*IAB-donor* and *IAB-node*), which are based on gNB-CU and gNB-DU entities with enhanced functionality.

An IAB-donor is the root of a logical multi-hop tree formed by IAB-nodes. Multiple IAB-donors that manage the respective trees may exist in an IAB-based RAN.

An IAB-node has a single parent node (the IAB-donor or another IAB-node) and none, one or multiple child IAB-nodes. The branches of the tree from leaf IAB-nodes up to the IAB-donor constitute the NR backhaul links. If there are feasible unused wireless links on the physical infrastructure, it is possible to explore alternative (redundant) paths to dynamically reconfigure the logical tree, under the control of the IAB-donor, if required (e.g. due to a backhaul link failure). In such case, some IAB-nodes may have to change their respective parent nodes.

An example of a simple IAB-based network (RAN and Core) is illustrated in Figure 8.

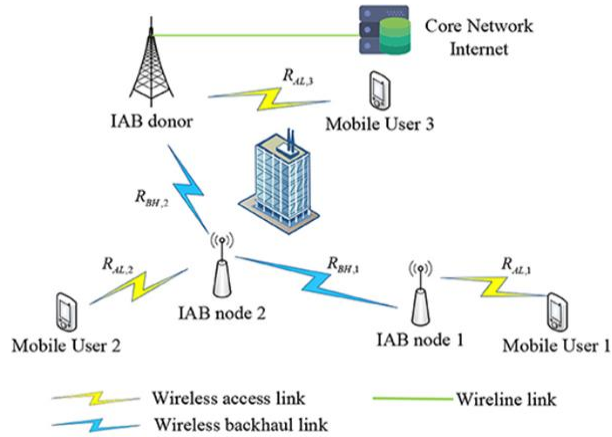

Figure 8. Example of IAB network [21].

The IAB-donor and the IAB-nodes also provide NR access links to UEs. All CP and UP user traffic is exchanged between UEs and the IAB-donor, which provides connectivity to the 5G Core Network.

Backhauling occurs when a UE is served by an IAB-node. In that case, UE traffic is carried by the backhaul links that form the path between the IAB-node and the IAB-donor; for a path with two or more hops, intermediate IAB-nodes simply act as relays.

### *IAB architecture and functional entities*

The IAB architecture defined in [16] is represented in Figure 9, which shows one IAB-donor, two cascaded IAB-nodes and the following interfaces: F1, backhaul NR Uu, NG and Xn-C. UEs and NR Uu access interfaces are not represented.

An IAB-donor is an enhanced gNB split into an *IAB-donor-CU* and one or multiple *IAB-donor-DUs*.

 An IAB-donor-CU controls each IAB-donor-DU via a direct (intra-donor) point-to-point F1interface

An IAB-donor-DU supports the NR Uu access interface with UEs and the NR Uu backhaul interface with child IAB-node(s), if any, and behaves as a gNB-DU on this interface.

An IAB-node is composed of two logical entities: an enhanced *Distributed Unit* (IAB-DU) and a new *Mobile Termination* (IAB-MT), which are designated, respectively, as gNB-DU and IAB-UE in [10].

An IAB-DU, similarly to an IAB-donor-DU, terminates the NR Uu access interface and the downstream NR Uu backhaul interface. It is controlled by the corresponding IAB-donor-CU via an indirect point-to-point F1 interface across backhaul link(s). This F1 interface carries CP and UP traffic of locally attached UEs, but is not involved in traffic relaying between backhaul links through the IAB-node

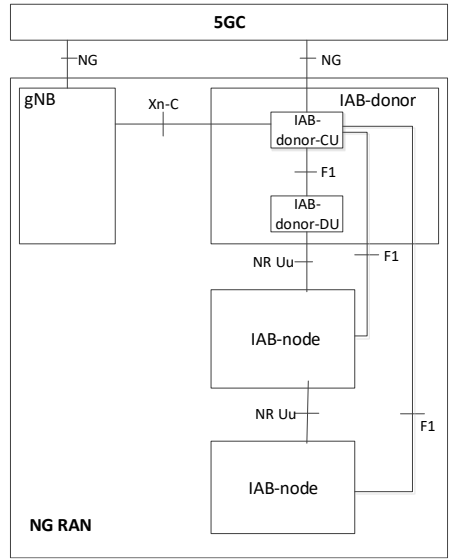

Figure 9. Overall architecture of IAB [16].

 The IAB-MT supports the NR Uu backhaul interface with the parent node; more specifically, the logical entity at the other endpoints of the backhaul link is an IAB-DU or an IAB-donor-DU.

An IAB-MT behaves as a UE and uses UE signalling procedures to connect to: 1) the gNB-DU entity of the parent node for access and backhauling, 2) the IAB-donor-CU via RRC for control of the access and backhaul link, and 3) the Core Network AMF via NAS for registration of the IAB-node and authorization to join the IAB network.

Internally to an IAB-node, CP and UP traffic is transferred, in both directions, between the IAB-DU, which terminates the access and the downstream backhaul interfaces, and the IAB-MT, which terminates the upstream backhaul interface. To deal with the two interfaces supported by the IAB-DU and the different procedures to execute in each case, the IAB-node must play two independent roles (*access node* and *relay node*), which may coexist.

As an *access node*, it transfers traffic between the access and the upstream backhaul links; in this case, the IAB-DU exchanges UE traffic with the IAB-donor-CU over the F1 interface, via the IAB-MT.

As a *relay node*, it transparently transfers traffic between the downstream and the upstream backhaul links; traffic is simply relayed between the IAB-DU and the IAB-MT.

### IAB protocol layers

In an IAB network, F1-C and F1-U traffic between an IAB-DU and the IAB-donor-CU is backhauled via the IAB-donor-DU and, possibly, some relay IAB-node(s). This requires extending and enhancing the F1 interface defined for the conventional CU/DU split gNB.

On the wireless backhaul (BH) links, the IP layer is carried over the *Backhaul Adaptation Protocol* (BAP) sublayer [22], which supports relaying over multiple hops. On each BH link, BAP PDUs are carried on BH RLC channels and may be mapped to multiple channels for traffic prioritization and QoS enforcement.

To illustrate the protocol stacks of all logical entities (IAB-donor-CU, IAB-donor-DU, IAB-MT and IAB-DU) and the relevant interfaces, a simple scenario with one IAB-donor and two IAB-nodes is considered. It allows distinguishing, in a clear way, the two roles played by IAB-nodes.

Control plane stacks for this scenario are shown in Figure 10.

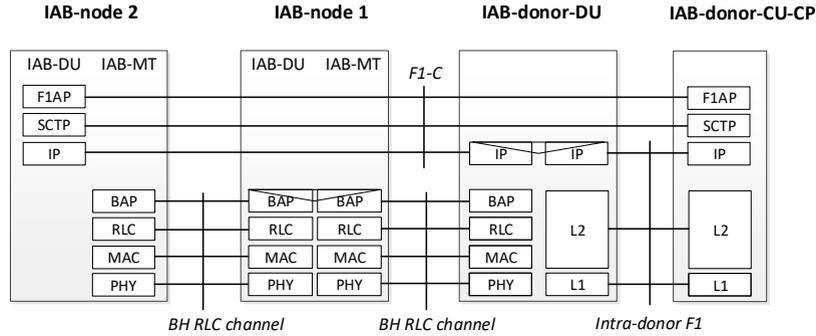

Figure 10. IAB protocol stack that supports F1-C [16].

IAB-node 2 acts as an access node. On the NR Uu access interface (not shown), IAB-DU 2 behaves as a gNB-DU, thus implementing the lower protocol layers (RLC, MAC and PHY). In addition, IAB-DU 2 terminates the F1-C interface protocols (F1AP, SCTP and IP) that run on top of the IAB-MT stack (BAP, RLC, MAC and PHY) on the upstream BH interface with IAB-node 1. The IAB-donor-CU terminates the higher protocol layers (RRC and PDCP), which are not represented; the respective PDUs exchanged with UEs are carried on F1AP messages between the F1-C endpoints.

IAB-node 1 acts as a relay node for IAB-node 2; thus, both IAB-DU 1 and IAB-MT 1 implement BAP on top of RLC. When serving locally attached UEs, IAB-node 1 implements the stacks and procedures as IAB-node 2, with its own F1-C interface.

User plane stacks for the same example are shown in Figure 11.

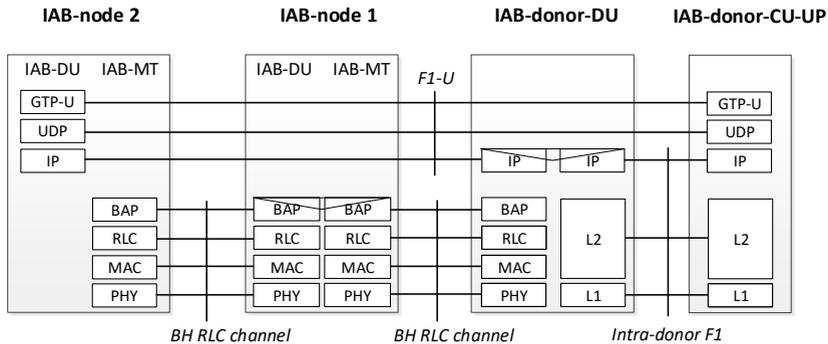

Figure 11. IAB protocol stack that supports F1-U [16].

Similar considerations apply, except that GTP-U, UDP and IP are the Transport Network Layer (TNL) protocols for the F1-U interface between IAB-DU 2 and the IAB-donor-CU. In addition, SDAP and PDCP protocols are executed between UEs and the IAB-donor-CU and the respective PDUs are carried by GTP-U.

To connect IAB-node 2 to IAB-node 1, as well as to the IAB-donor and the Core Network, IAB-MT 2 behaves as a UE and reuses UE procedures (with enhancements), as explained. In this context, IAB-MT 2 must establish *Signalling Radio Bearers* (SRBs) at the PDCP layer to exchange RRC and NAS traffic with the IAB-donor-CU and the 5G Core AMF, respectively. Acting as a UE has three implications:

- IAB-MT 2 implements a complete NR Uu Control Plane stack, with NAS on top of the NR Uu protocol layers (RRC down to PHY).
- It communicates with IAB-DU 1 on the NR Uu access link, both implementing RLC, MAC and PHY protocols.

- It relies on the F1-C interface established by IAB-DU 1 to exchange RRC messages with the IAB-donor-CU (that terminates RRC and PDCP) and NAS messages with AMF. NAS messages are carried by RRC on F1-C and by the *NG Application Protocol* (NGAP) on the NG N2 interface between the IAB-donor CU and AMF.

Simplified protocol stacks for handling SRBs are represented in Figure 12, which only shows the peering of the IAB-MT 2 protocol layers. The protocol stacks on the F1-C interface between IAB-DU 1, which treats IAB-MT 2 as a UE, and the IAB-donor-CU are not represented as well.

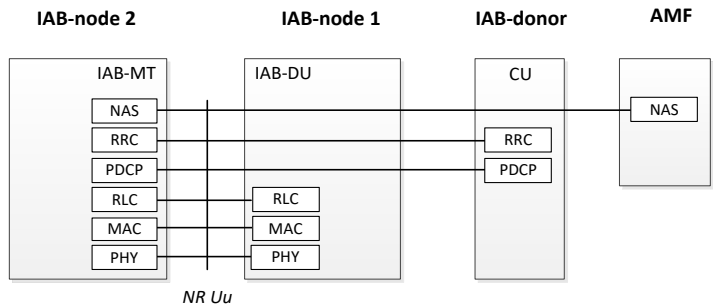

Figure 12. Protocol stack for the support of IAB-MT's RRC and NAS connections [11].

The integration of an IAB-node in an IAB-based RAN consists of three steps: 1) IAB-MT set-up, 2) BH RLC channel establishment and routing update, and 3) IAB-DU set-up. Details are given in [16].

## 2.4    IABIAB-based Mobile Base Station Relay (MBSR)

A *Mobile Base Station Relay* (MBSR) is defined in [23] as a mobile base station acting as a relay between a UE and the 5G network. An MBSR is mounted on a moving vehicle and serves UEs that can be located inside or outside the vehicle (or entering or leaving the vehicle).

Support for MBSR operation is described in [10]. An MBSR operates as an IAB-node in an IAB network, with mobility support when integrated with the serving PLMN, with the following particularities:

- It must be at one hop from the IAB-donor, that is, it cannot be attached to an IAB-node.
- As an IAB-node, the NR Uu interface is used for the UE access link and for the backhaul link to the IAB-donor, unlike a UE relay, which uses a PC5-based link to provide an indirect connection to remote UEs [24].

Roaming of MBSRs is possible if there is a roaming agreement for MBSR operation between the *Visited PLMN* (VPLMN) and the *Home PLMN* (HPLMN). The concept of *Serving Network* is used to refer to the IAB network where the MBSR is integrated with the respective IAB-donor; thus, the *Serving Network* is either the HPLMN or, when roaming, the VPLMN.

### *MBSR authorization and configuration*

The MBSR subscription information is stored in the HPLMN and indicates whether it is authorized to operate as an MBSR, and the corresponding location and time-periods.

To register and request to operate as an MBSR on the serving PLMN, the MBSR IAB-MT provides a *mobile IAB-indication* to the IAB-donor-CU when the RRC connection is established; then, the IAB-donor-CU selects an AMF that supports the mobility of IAB-nodes. The AMF authorizes the MBSR based on the subscription information and provides MBSR *authorized indication* to the MBSR (via NAS) and to the NG-RAN (via NGAP), as described in [25].

For non-roaming MBSRs, the operator's local policy can be considered for MBSR authorization, e.g. based on the network state or on a limit on the number of MBSRs operating in a certain area.

To operate as a mobile IAB-node, an MBSR receives configuration from the *Operation and Maintenance* (OAM) system of the serving PLMN. The MBSR IAB-MT establishes a secure and trusted connection to the OAM server only if it is authorized to operate as an MBSR in the serving PLMN. An MBSR can be preconfigured with UE policy or provisioned using existing UE Policy mechanisms [26].

After the IAB-MT is registered in the 5G System, further mobility procedures can be executed to change the IAB-donor-DU or the IAB-donor-CU, as specified in [16].

### Mobility support of UEs served by an MBSR

Three scenarios and the respective procedures are described in [10]:

- *UE mobility between a fixed cell and an MBSR cell* – the procedure of Inter-gNB-DU Mobility defined in [16] or the handover procedure using the Xn/N2 reference points defined in [25] can be used.
- *UE mobility between MBSR cells* – UEs use the same procedures as in the previous case, under similar conditions, to handle mobility between MBSR cells.
- *UE mobility when moving together with an MBSR cell* – the IAB-donor-CU or the OAM server may configure the *Tracking Area Code* (TAC) broadcasted by the MBSR cell(s).

### Control of UE access to an MBSR

A *Closed Access Group* (CAG) identifier is used to control UE access via an MBSR and the existing CAG mechanism defined in [10] can be used for managing such accesses.

When an MBSR is allowed to operate as MBSR for a PLMN, it is configured with a *CAG* identifier (CAG-ID), unique in the scope of the PLMN, during communication with the serving PLMN OAM or preconfigured. If the MBSR is preconfigured with the list of PLMNs in which it is allowed to operate, the corresponding CAG-ID per PLMN is also configured in the MBSR.

Time duration restriction may be provided together with the CAG-ID for the MBSR(s) that can be accessed by a UE.

## 3    5G Mobile Cell design – methodology and models

### 3.1    Rationale and methodology

The 5G Mobile Cell (MC) under development is intended to provide subscribers of a 5G network (a PLMN or an SNPN) with access to its core services, when UE access through the stationary RAN gNBs is not possible (e.g. when not in radio range) or has poor quality. This is the *Home Network* (HN) for the MC and the subscribed UEs.

In this paper, we assume that the MC is not shared by the HN operator with other operators and that UEs with subscriptions on other networks are treated as when roaming on a stationary RAN, thus having no impact on this analysis.

Before starting the design of the MC, a high-level analysis of possible alternative architectures was carried out.

Initially, an IAB-based solution emerged as promising and interesting for several reasons:

- The IAB architecture specification was stable and well documented, as discussed in Section 2.3.
- The specification of a *Mobile Base Station Relay* (MBSR) as a *mobile IAB-node* was underway and is now complete, as described in Section 2.4.
- Integrating the MC as a mobile IAB-node is less complex than adding a fixed IAB-node, since the MC is a leaf IAB-node (only serving UEs) at one hop from the IAB-donor (thus, with no parent IAB-node), thus allowing seamless integration of the access and backhaul links on the IAB RAN.

- It would offer new research opportunities leveraging existing knowledge of 5G and Wi-Fi networks, namely in system integration and testing, and simulation tools (e.g. ns-3).

For a proof-of-concept, the MC should be integrated with a real IAB-network, which proved to be not possible. For laboratory testing, functional validation, and performance evaluation, developing an IAB-donor would be sufficient, but the extra effort required was not worth it at that time.

Thus, we decided to explore an alternative architecture for the design, development and testing of an MC prototype in the short-term, and favoured a modular solution such that it might be possible, at a later stage, to adapt and enhance some components and reuse them to build an IAB MBSR, as initially intended.

Two aspects were analysed in order to identify possible solutions:

- Which gNB logical entities should be implemented in the MC; in other words, how to split gNB functions between the mobile platform and the fixed infrastructure.
- How to provide connectivity between the mobile platform and the fixed infrastructure needed to support MC operation.

## 3.2    Mobile Cell prototype – high-level models

Two architectures were selected for the MC prototype:

- The MC as a *mobile gNB* – for simplicity, we assume a *non-split CU/DU* architecture; however, the internal organization of the gNB entities is not constrained, since it is not visible to other systems.
- The MC as a *mobile gNB-DU relay* – it includes a gNB-DU and respective RU, controlled by a gNB-CU located on the fixed infrastructure; this implies a *CU/DU split 2* architecture and, thus, support of the F1 interface.

The interest of the second solution is twofold: 1) since the MC includes a DU module and terminates the F1 interface, it may be enhanced and converted into an IAB MBSR, and 2) the CU and DU modules developed for this solution may be reused and integrated to build a *mobile gNB*.

Independently of the physical split of the logical gNB entities, it is necessary to establish IP connectivity, over a wireless channel, between the MC and a target 5G network that includes RAN stationary nodes and 5G Core functions.

An obvious alternative to the *integrated model* is to use a 5G *Overlay Network* (ON) for this purpose. It is worth mentioning that 3GPP adopted the *overlay model* to support two services specified in [10]: UE access to SNPN service via a PLMN and vice-versa.

The ON provides IP connectivity across its RAN and Core by means of a *PDU Session* set up by the MC, and IP routing to external *Data Networks*. Multiple *PDU Sessions* may be established, if needed.

The natural solution is to connect the MC to its *Home Network* (HN) and integrate it as a mobile RAN node, as outlined hereafter. In this way, the MC virtually operates in its HN environment; thus, the MC and UEs that have subscriptions on this network and use the MC to access its service are never roaming.

To analyse real MC operation scenarios, we must look at the ON and the HN not only as physical networks but also from the point of view of their logical roles. In this context, the ON emulates a backhaul link, thus playing the role of *Transport Network* (TN), while the HN plays the role of *Serving Network* (SN). Thus, TN and SN must be seen as logical networks, abstracting from the physical network(s) that play these roles. To clarify this distinction in specific scenarios, we assume that the MC's HN is an SNPN.

When the MC operates within radio range of stationary gNB(s) of its HN (the SNPN), the ON is, by default, the SNPN (acting as TN), which also acts as SN; thus, both roles are played by a single network (the SNPN)

When the MC is outside SNPN coverage, the ON must be another network, e.g. a neighbour PLMN, which plays the TN role only, while the SNPN (the MC's HN) still plays the SN role. The SN and TN roles are played by different networks.

As an extension of the first scenario, a PLMN may be also used as ON to improve service, either when the quality of the SNPN RAN access link degrades and handover to the PLMN is possible or to distribute MC traffic between the two overlay networks (SNPN and PLMN).

The MC procedures required to establish IP connectivity with the MC's HN are executed by a *Mobile Termination* entity (MC-MT) that terminates the ON backhaul link.

First, a 5G channel is setup between the MC-MT, acting as a UE, and a gNB on the ON RAN (ON-gNB); the MC-MT is agnostic to the architecture of the ON-gNB(s).

Using standard UE procedures, the MC-MT must register and authenticate with an AMF on the ON Core (ON-AMF). When the ON is the MC's HN, the MC-MT uses its HN subscription; otherwise, it either needs a subscription on the ON or a roaming agreement with the ON to use the HN subscription.

Once the MC is authorized to operate, a *PDU Session* is setup between the MC-MT and an UPF on the ON Core (ON-UPF). Then, this *PDU Session* is used to transparently carry through the ON all CP and UP traffic exchanged between UEs and their HN.

The two MC solutions based on the *overlay model* differ on the entities at the endpoints of the MC-MT *PDU Session*, as follows.

An MC implemented as a *mobile gNB* consists of CU, DU, and RU entities (typically based on a *non-split CU/DU* architecture) and an MC-MT on board a mobile platform. IP connectivity over the ON enables the communication between the MC and 5G Core Network functions on the fixed HN infrastructure, as depicted in Figure 13 [9].

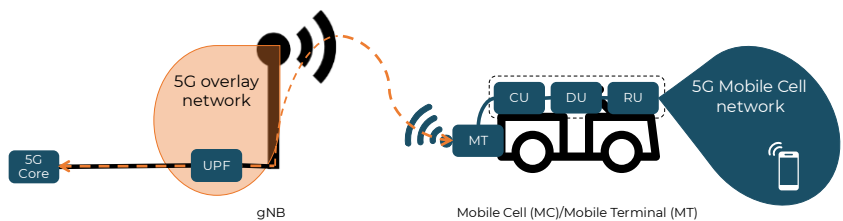

Figure 13. MC based on CU, DU and RU on a mobile platform.

An MC implemented as a *mobile gNB-DU relay* consists of an MC-DU (and respective RU) and an MC-MT on board a mobile platform. The communication between the MC-DU and an MC-CU on the fixed HN infrastructure, over an F1 interface, is enabled by IP connectivity setup across the ON. The MC-CU communicates with 5G Core Network functions typically using wired technologies. This is illustrated in Figure 14 [9].

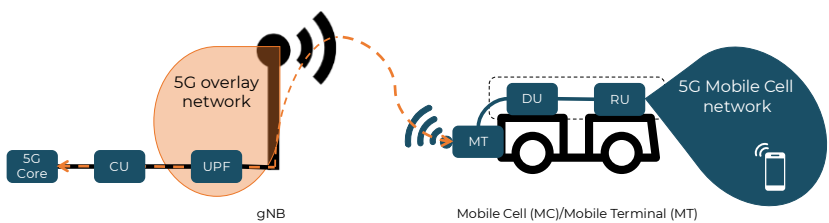

Figure 14. MC based on DU and RU on a mobile platform and CU on the fixed infrastructure

For a complete assessment of these solutions, we also consider in the following analysis the IAB-based solution: the *MC as a mobile IAB-node*, i.e. an IAB MBSR.

### 3.3 Mapping to 3GPP reference models

The concept of a *Mobile Base Station Relay* (MBSR) based on the IAB architecture was introduced by 3GPP in Release 18 and the first specification appeared in Version 18.0.0 (2022-12) of TS 23.501 [10], which was updated and enhanced in subsequent versions. The IAB MBSR is the reference for the *MC as a mobile IAB-node* solution.

In the context of the Release 19 work plan, a *Study Item* (SI) on *Additional Topological Enhancements for NR* was proposed and the respective description [27] was approved in March 2024. This SI introduced the concept of *Wireless Access Backhaul* (WAB) and proposes as the main objective to study the architecture and protocol stack for supporting a *"gNB with MT function providing PDU session backhaul"*.

A new Technical Report (TR 38.799) [28] was created to keep track of progress on this SI with new content, after approval by the 3GPP *Radio Access Network Technical Specification Group*. Version 19.0.0 (2024-09) is the first formal document with Release 19 numbering, enhancing Version 1.0.0 (2024-06).

The concept and main principles adopted for the design of the *MC as a mobile gNB* coincide with the requirements set for WAB, as well as with the architecture and supporting protocol stacks [28]. For characterization and comparison purposes, we consider *MC as a mobile gNB* equivalent to WAB, the current 3GPP reference for a solution based on the *overlay model*.

The *MC as a mobile gNB-DU relay* solution shares characteristics with WAB (*overlay model*) and IAB MBSR (CU/DU split architecture and, thus, support of the F1 interface); although not standardized, its components may be adapted and/or enhanced to build WAB and IAB compliant solutions.

Based on the high-level models (*overlay* and *integrated*), the gNB functional split options and the end-to-end scenario in which an MC operates, the protocol stacks relative to the three MC solutions under study will be derived and discussed in Section 4.

These solutions will be analysed and compared using different criteria in Section 5.

## 4 MC architectures – complete protocol stacks

As explained, the two MC solutions based on the *overlay model* require the establishment of a *PDU Session* by an *Overlay Network* (ON), between the MC-MT and an UPF on the ON (ON-UPF).

### 4.1 Establishment of the PDU Session on the overlay network

To set up the *PDU Session*, the MC-MT must first authenticate and be authorized to access the ON, as explained in Section 3.2. For this purpose, the MC-MT selects a gNB on the ON (ON-gNB) and uses standard UE procedures on the Control Plane (CP) to register (on behalf of the MC) on an ON-AMF selected by the ON-gNB. The ON-AMF manages MC access and mobility through the MC-MT, which is treated as a UE. Once registered and authorized, the MC-MT uses standard UE procedures on the User Plane (UP) to establish an MC-MT *PDU Session*. The respective CP and UP protocol stacks are represented in Figure 15.

The MC-MT communicates with the ON-gNB over the NR Uu interface [10], using the CP and UP standard protocols illustrated in Figure 4, and with the ON-AMF, using CP NAS signalling procedures.

On the CP, the ON-gNB communicates with the ON-AMF over the standard N2 interface, using NGAP over SCTP, IP and lower layer (L2 and L1) protocols.

On the UP, the ON-gNB communicates with the ON-UPF over the N3 standard interface, using GTP-U over UDP, IP and lower layer protocols.

The MC-MT *PDU Session*, which terminates on the ON-UPF, is subsequently used to carry CP and UP user traffic across the ON. Thus, the UP stacks in the figure are a building block common to both ON-based MC architectures and will be reused in the respective diagrams.

The MC is unaware of the ON-gNB architecture: the MC-MT may attach, via the NR Uu interface, to a monolithic gNB, to a gNB-DU of a CU/DU split gNB or, in an IAB-based ON, to an IAB-donor-DU.

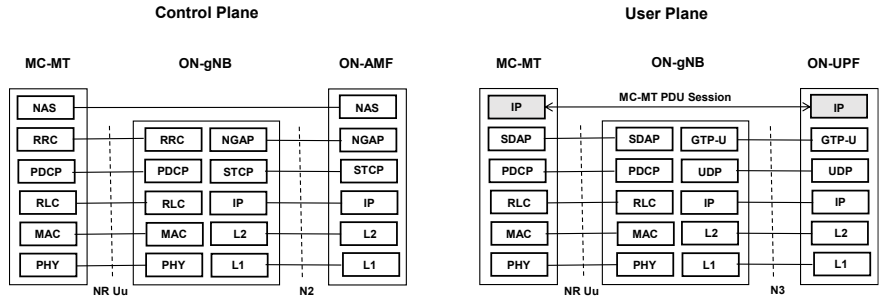

Figure 15. CP and UP protocol stacks for establishing and using the MC-MT PDU Session.

The complete CP and UP protocol stack for both overlay solutions, which include the MC and ON stacks, as well as the UE and *Home Network* (HN) Core stacks, will be discussed next. For comparison purposes, we also provide the CP stacks for an MC implemented as *mobile IAB-node* at one hop from an IAB-donor.

## 4.2 MC as a mobile gNB

The MC-MT, the ON-gNB and the ON-UPF protocol stacks, as building blocks, may be abstracted and seen as emulating a backhaul link between the MC-gNB and the HN Core functions. They are shaded to highlight the IP-based transport mechanism.

As explained, we only consider UEs whose HN is the MC's HN. Thus, UEs use the MC-gNB to communicate, via the ON, with HN Core functions, namely AMF on CP (HN-AMF) and UPF on UP (HN-UPF).

The communication between the ON-UPF and HN Core functions takes place over an external IP network, not represented in the following figures.

The CP stacks shown in Figure 16 comply with those that apply to UE, gNB and AMF interfaces. The UE and gBN stacks on the NR Uu access interface and the NAS signalling protocol between the UE and the HN-AMF are the same as on the left side of Figure 4 (and Figure 15).

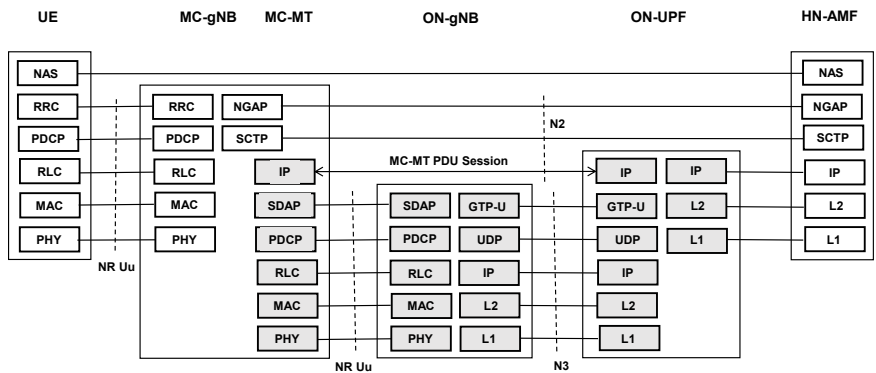

Figure 16. MC as a mobile gNB – CP protocol stacks.

The MC-gNB and the HN-AMF communicate over the standard N2 interface, using NGAP over SCTP; these protocols are carried on the emulated backhaul link (over the MC-MT *PDU Session*).

The HN-AMF manages UE access and mobility. When the ON role is played by the MC's HN, the ON-AMF and the HN-AMF may be the same, if the AMF is configured to support procedures specific to the MC-MT (e.g. registration and authorization or handover).

The CP signalling procedures are used to set up a UE *PDU Session* on the UP, whose protocol stacks are shown in Figure 17. The UE and gNB stacks on the NR Uu access interface are the same as on the right side of Figure 4 (and Figure 15).

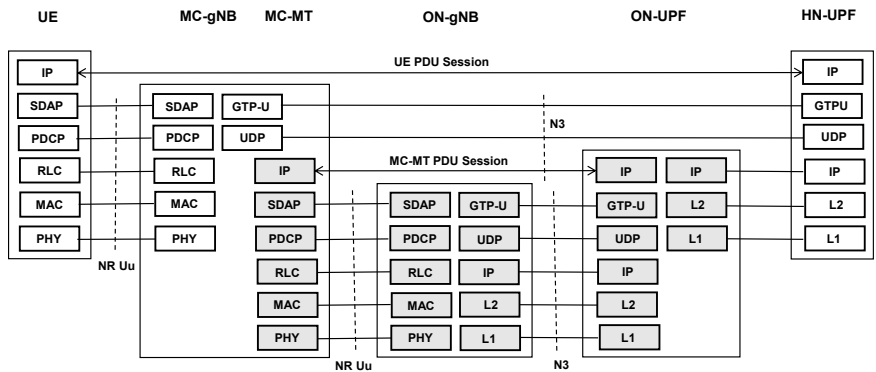

Figure 17. MC as a mobile gNB – UP protocol stacks.

The MC-gNB and the HN-UPF communicate over the standard N3 interface, using GTP-U over UDP; these protocols are also carried on the emulated backhaul link (over the MC-MT *PDU Session*).

The UE *PDU Session* between a UE and an HN-UPF is used to carry IP user traffic within the HN (represented by the MC-gNB, the HN-AMF and the HN-UPF).

When the ON role is played by the MC's HN, the ON-UPF and the HN-UPF may be the same.

## 4.3    MC as a mobile gNB-DU relay

The MC-gNB is split into an MC-DU (coupled with an MC-MT) on a mobile platform and an MC-CU on the HN fixed infrastructure, which communicate over the F1 interface (F1-C and F1-U).

The CP protocol stacks are illustrated in Figure 18.

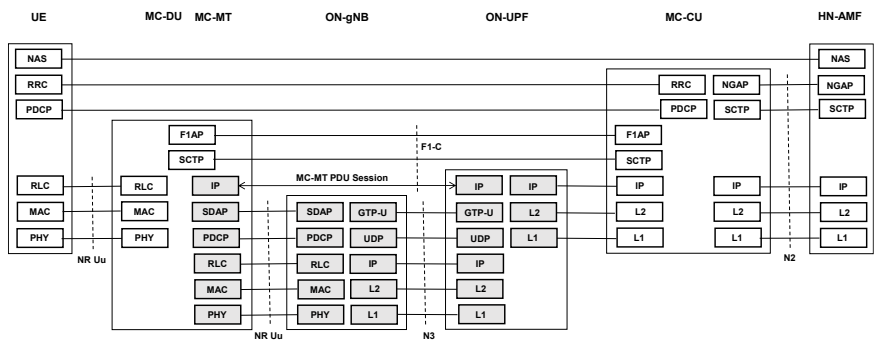

Figure 18. MC as a mobile gNB-DU relay – CP protocol stacks.

The UE protocol stack does not change (the UE is unaware of the MC architecture), including NAS, which is terminated at the HN-AMF. RRC and PDCP are handled by the MC-CU; placing it away from the mobile platform has three implications:

- The MC-DU only terminates lower protocol layers (RLC, MAC and PHY) on the NR Uu interface.
- RRC and PDCP are carried on F1AP messages on the F1-C interface (between MC-DU and MC-CU, over the MC-MT *PDU Session*).
- The N2 interface between the MC-CU and the HN-AMF is supported on wired media (as in stationary gNBs) and not across the ON.

Besides terminating F1AP and SCTP on the F1-C interface, as well as RRC and PDCP, the MC-CU supports N2 interface protocols (NGAP and UDP) with the HN-AMF.

The UP stacks represented in Figure 19 are derived from the CP stacks, with the appropriate changes:

- SDAP and PDCP on the UE and on the MC-CU (NR Uu interface).
- GTP-U and UDP on the MC-DU and on the MC-CU (F1-U interface).
- GTP-U and UDP on the MC-CU and on the HN-UPF (N3 interface).

Similarly, the UE *PDU Session* is established between the UE and the HN-UPF.

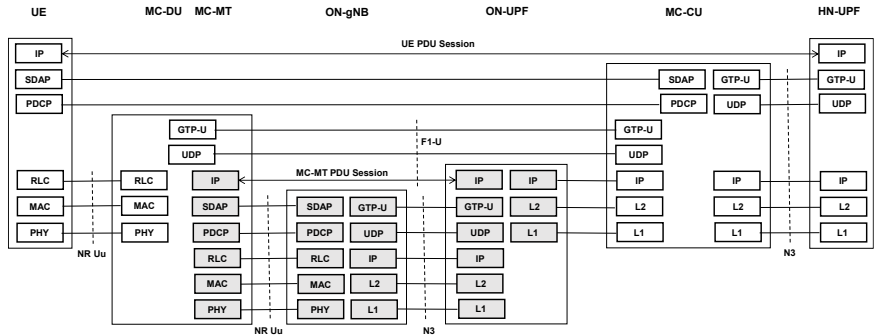

Figure 19. MC as a mobile gNB-DU relay – UP protocol stacks.

For the MC prototype and proof of concept, the MC-CU has been developed as an autonomous system. However, in case HN gNBs are based on a CU/DU split architecture, a gNB-CU may be used to control the MC-DU like any attached gNB-DU.

## 4.4 MC as a mobile IAB-node

We use the same notation for the MC entities as in the overlay cases. Since the MC is a mobile IAB-node, the MC-DU and the MC-MT represent an IAB-DU and an IAB-MT, respectively.

Since in the IAB-based solution there is a single integrated serving network (the home or a visited network), we use a different notation for the core functions: MT-AMF, UE-AMF and UE-UPF.

To integrate the MC into the IAB network, the MC-MT acts as a UE and executes RRC and NAS signalling procedures, respectively with the IAB-donor-CU and the MT-AMF, which manages MC access and mobility (via the MC-MT). The protocol stacks for the MC-MT, the IAB-donor-DU, the IAB-donor-CU (including the intra-donor F1-C interface) and the MC-MT are illustrated in Figure 20, which is a detailed version of Figure 12. Since the MC-MT acts as a UE at one hop from the IAB-donor, backhaul relaying is not needed; thus, BAP is not present.

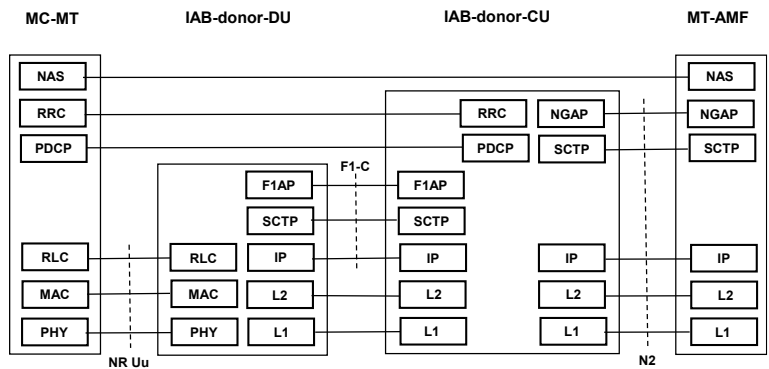

Figure 20. MC as a mobile IAB-node – MC-MT signalling stacks.

The full CP protocol stacks required for UE signalling with the IAB-donor-CU and the UE-AMF, which manages UE access and mobility, are illustrated in Figure 21. RRC and PDCP are carried on F1AP messages over the F1-C interface between the MC-DU, which provides the UE access link on the NR Uu interface, and the IAB-donor-CU. The MT-AMF (in Figure 20) and the UE-AMF may be the same, since they belong to the same Core Network that serves the IAB RAN.

For comparison (especially with the CP stacks for the *MC as a mobile gNB-DU relay*, in Figure 18), the MC-MT and IAB-donor-DU stacks are also shaded.

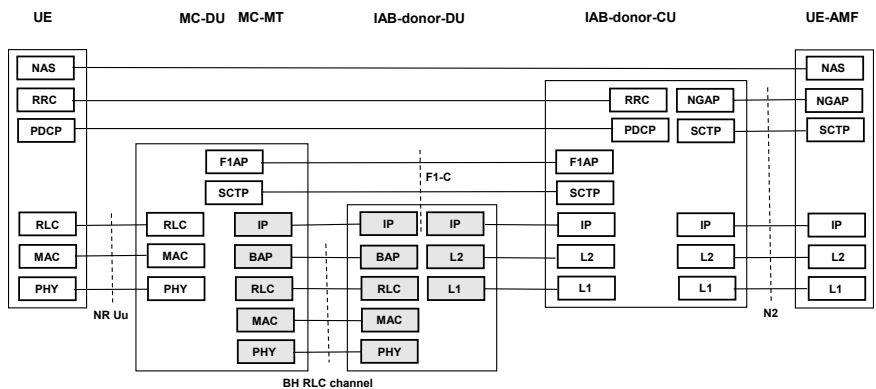

Figure 21. MC as a mobile IAB-node – CP protocol stacks.

Unlike the overlay model, no UPF is involved in CP signalling procedures for both the MC-MT (Figure 20) and the UE (Figure 21).

The UP stacks may be easily derived but are not represented, since the CP stacks are sufficient for comparison purposes. At the IP level in the UP , the UE *PDU Session* terminates on a UE-UPF.

# 5    Analysis and comparison of Mobile Cell solutions

First, the comparison of MC solutions is based on the respective protocol stacks, using the Control Plane as reference (when relevant, User Plane protocols are invoked).

Since WAB and IAB are the 3GPP reference architectures for, respectively, the overlay and the integrated models, we first compare *WAB / MC as a  mobile gNB* (Figure 16) with *IAB MBSR / MC as a mobile IAB-node* (Figure 21). These figures put into evidence two fundamental architectural differences: 1) which functions reside in the mobile platform, and 2) the IP *Transport Network* (TN) specific to each backhauling method. The protocol stacks that apply to each TN scheme are shaded in the respective figures and used for comparison.

In WAB, the MC behaves as a full gNB (regardless of the internal organization of its functional entities):

- The MC-gNB terminates RRC/PDCP on the CP NR Uu interface and NGAP/SCTP on the CP N2 interface (SDAP/PDCP on the UP NR Uu interface and GTP-U/UDP on the UP N3 interface).
- The MC-MT, acting as a UE, terminates a full UP protocol stack on the NR Uu interface to carry IP packets; thus, it is possible to reuse UE standard components and procedures.

The TN introduces an ON-gNB and an ON-UPF on the CP path (Figure 16) between UEs and an HN-AMF, and on the UP path (Figure 17) between UEs and an HN-UP. Even when the HN plays the role of TN and Core functions may be shared, there are two gNBs on the CP/UP paths (ON-gNB and MC-gNB) and ON and HN instances of Core functions: ON-UPF (CP/UP), HN-AMF (CP) and HN-UPF (UP), while an ON-AMF (Figure 15) is first invoked to set up the MC-MT *PDU Session*.

In IAB, the MC is simply a mobile IAB-node:

- The MC-DU does not terminate RRC and NGAP on CP (SDAP and GTP-U on UP), which are delegated to the IAB-donor-CU. Instead, it must terminate the F1 interface (F1AP/SCTP on CP and GTP-U/UDP on UP); despite its complexity, F1AP is a standard for the CU/DU split architecture.

- The MC-MT, first acting as a UE, registers the MC-DU as a mobile IAB-node with the IAB-donor-CU (Figure 20) and then, it uses the BH RLC channel to carry IP packets over BAP: although standardized, these procedures are specific of the IAB architecture

IP connectivity between the MC-MT and the IAB-donor-CU is provided by the IAB-donor-DU, that is, the IP transport mechanism is natively supported by the IAB *Serving Network* (HPLMN or VPLMN). There is a single Core: an MT-AMF is invoked by the MC-MT to register the MC-DU (Figure 20) and during MC operation the UE communicates with the IAB-donor-CU and Core functions (AMF and UPF).

When comparing both solutions, we must analyse the complexity (and processing overhead) of the mobile system and the factors that affect the performance of the overall solution.

The MC-gNB (WAB) is more complex than the MC-DU (IAB), especially in the CP, due to RRC/NGAP processing, when compared with F1AP procedures (F1 interface)

Considering the overall solution, the processing and transmission overheads introduced by the TN are considerably higher in the WAB case; however, this difference is partially compensated when taking into account the IAB-specific processing and transmission overhead introduced by the IAB-donor (termination of the F1 interface and two additional internal IP hops).

This qualitative analysis based on complexity and the impact of overheads on performance should be confirmed with a quantitative evaluation by simulation and in field trials. However, it is likely that these criteria alone will not be decisive to choose one solution suitable for all scenarios. Rather, a decision must take into account other factors and how they are valued in each specific case.

For a more comprehensive characterization of WAB behaviour (and comparison with IAB using other criteria), we take as reference the current contributions to the WAB specification. A few issues are still under study, aimed at designing solutions that comply with WAB requirements and assumptions, namely mobility of WAB nodes and adoption of the overlay model.

In WAB, the mobility of the gNB-CU entity (involving RRC/NGAP/XnAP) implies that neighbour relations with other gNBs (stationary or mobile) are no more static. Thus, it is necessary to study the impact on *Automatic Neighbour Relation* (ANR), *inter-gNB handover* and *Self Organizing Network* (SON) functionality, and identify the need to enhance inter-gNB and gNB to Core Network signalling to support dynamic inter-gNB interactions and the respective procedures.

Unlike IAB, WAB does not yet support end-to-end QoS; thus, it must be studied how to enhance QoS across the overlay (backhaul) network. Similarly, it is necessary to study which signalling enhancements are required to authorize the operation of mobile WAB nodes.

To complete this analysis focused on WAB and IAB, we consider the *MC as a mobile gNB-DU relay* solution and the respective CP stacks (Figure 18) for comparison.

This is, in its objectives, an experimental solution. From the previous analysis, we may conclude that it is clearly outperformed by the WAB and IAB solutions. Nevertheless, it allows assessing in the same platform the impact on performance of the overlay TN and the F1 interface; moreover, its components may be adapted or enhanced to build the WAB and IAB solutions.

When comparing with WAB, the only difference resides on CU/DU splitting and moving the MC-CU to the fixed infrastructure. This requires supporting the F1 interface over the MC-MT *PDU Session*. Since the MC-CU must terminate RRC with UEs and NGAP with HN-AMF in CP (SDAP and GTP-U in UP), similarly to a full gNB, the extra complexity is due to the F1 stacks on the MC-DU and the MC-CU and, as a consequence, two IP hops are added by the MC-CU.

Development of the MC-CU may be avoided when the HN RAN includes CU/DU split gNB(s), in which case the MC-DU may be controlled by an HN gNB-CU; otherwise, a special purpose MC-CU must be developed and deployed on the HN.

This architecture introduces a potential problem that is not present in WAB or IAB. During UE initial attachment (or reattachment), the UE sends an *RRC Connection Request* message to the selected gNB and expects to receive as response an *RRC Setup* or an *RR Reject* message [29]. This message exchange is protected by a *Contention Resolution Timer* (CRT), whose maximum configurable value is 64 ms. Thus, this is the maximum acceptable r*ound-trip-time* (RTT) between the UE and the MC-CU that avoids CRT expiry.

In comparison with WAB, RTT increases due to F1 processing on the MC-DU and the MC-CU, and to the two-way delay on the overlay network (ON-gNB and ON-UPF). CRT expiry, which prevents successful UE attachment, was observed during preliminary tests, mainly due to processing limitations; the problem was solved by increasing the CRT value.

The comparison with IAB MBSR is simple. Both architectures are based on CU/DU split, thus requiring support of the F1 interface, while the MC-CU and the IAB-donor-CU are functionally similar; thus, the main difference is due to the backhauling method. The overlay model introduces higher processing and transmission overheads (ON-gNB and ON-UPF) than the integrated model (IAB-donor-DU), as discussed.

Finally, for a comprehensive comparison of the three outlined solutions, a summary of their main attributes and properties, which allows identifying similarities and differences, is provided in Table 1.

Table 1. Comparison of Mobile Cell solutions.

| MC as a mobile gNB | MC as a mobile gNB-DU relay | MC as a mobile IAB-node |
|---|---|---|
| *Overlay model* – mobile gNB (WAB node) | *Overlay model* – gNB-DU relay. | *Integrated model* – IAB-DU relay (MBSR). |
| The MC hosts a gNB (MC-gNB) and an MC-MT. | The MC hosts an MC-DU and an MC-MT (CU/DU split gNB) | The MBSR hosts an IAB-DU and an IAB-MT (IAB architecture). |
| The overlay network introduces processing and transmission overheads | The overlay network and the F1 interface introduce processing and transmission overheads, with risk of CRT expiry. | The F1 interface introduces processing and transmission overheads |
| It is possible to deploy an UPF on the MC and provide local services to UEs, as it hosts the MC-CU. | It is not possible to deploy an HN-UPF on the MC and provide local services to UEs, as it does not host the MC-CU. | It is not possible to deploy an UPF on the MBSR and provide local services to UEs, as the CU entity resides in the IAB-donor. |
| MC-CU mobility creates dynamic relations with neighbour gNBs, which may require enhancing inter-gNB and gNB-to-Core Network signalling. | Signalling enhancements are not required, since the MC-CU is stationary and relations with neighbour gNBs remain static | An IAB-donor-CU is stationary; UE mobility may be supported by existing inter-gNB procedures. |
| Since the overlay model requires setting up an MC-MT *PDU Session*, there is an ON-UPF on the UE control (CP) and data (UP) paths | | There is no UPF within the IAB RAN, since relaying is performed at Layer 2 (BAP). |
| The MC is agnostic to the architecture of ON-gNBs. | | The IAB-DU is controlled by an IAB-donor-CU; thus, it can only operate on IAB networks. |
| Since the *Serving Network* is the MC's HN, no roaming agreement is required by UEs with subscriptions on this network | | A roaming agreement is required between the HPLMN and a VPLMN, when the VPLMN is the *Serving Network*. |
| The MC-MT uses the standard CP stack on the NR Uu interface with the ON-gNB to set-up the MC-MT *PDU Session* and the UP stack to transfer UE traffic (CP and UP). | | The IAB-MT has an IAB specific Layer 2 stack to exchange traffic with the IAB-donor-DU. |
| End-to-end QoS is not yet supported; thus, it is necessary to study how to enhance QoS across the ON. | | The IAB architecture supports end-to-end QoS. |
| Signalling enhancements for authorizing MC operation are also required. | | Signalling enhancements for IAB MBSR operation are specified. |

The content of this table is based on the respective architectures and other aspects derived from requirements and assumptions in 3GPP specifications (IAB and WAB).

Since the impact of processing and transmission overheads on performance has already been discussed, we focus on the other comparison criteria and start by highlighting the strongest points of the WAB-based solution.

The possibility of deploying an UPF on the MC is a sole property of WAB. This allows offering on-board services to UEs, which may be an advantage in some use cases (and a weakness of the other two solutions).

The overlay solutions share two relevant properties: independence of the architecture of the overlay (backhaul) gNBs and no need of roaming agreements for UEs with subscriptions on the MC's HN. This means that there is no restriction to extending 5G coverage and assure service continuity to such UEs when the MC moves away from the HN, except in areas where there is no PLMN coverage or PLMNs in which the MC is authorized to operate (no MC-MT subscription or lack of roaming agreement). UEs without subscriptions on the MC's HN are treated in the same way as when accessing through a stationary gNB, which depends on a roaming agreement.

Conversely, an IAB MBSR can only attach to an IAB network, i.e. to an IAB-donor, and, when roaming, an agreement with a visited IAB network must be in place. Thus, when moving away from the IAB HPLMN, extending 5G coverage and assuring service continuity to UEs not only depends on the availability of an IAB VPLMN but also on a roaming agreement with that network; this may severely restrict the areas in which the MC (IAB MBSR) may operate.

On the other hand, some WAB weaknesses have been identified and deserve further study to overcome or attenuate them.

First, the mobility of WAB nodes introduces dynamic relations with neighbour gNBs, which do not occur in stationary RANs. Solving this problem, which is exclusive to WAB, probably requires signalling enhancements yet to be studied. This is not a problem in IAB.

Another issue is the need to support QoS across the overlay network, which may require signalling enhancements as well; on the contrary, end-to-end QoS is supported in IAB.

Finally, the authorization issue is, probably, less problematic. Although signalling enhancements may still be required, authorization and configuration procedures assisted by an OAM server, similar to those specified for IAB, may be used.

In what concerns IAB MBSRs, a conclusion is due: the high level of integration achieved on a single serving network, with QoS support, may be outweighed by the weaknesses or limitations abovementioned, which may not recommend or even preclude its adoption in some use cases.

# 6    Mobile Cell development

In order to validate the concept of Mobile Cell and assess the performance of an MC as a *mobile gNB-DU relay* based on the overlay model, we employed the model proposed in [30]. It includes an F1 wireless interface emulator and the capability of inducing different path loss attenuations on the wireless backhaul link between an ON-gNB and an MC, as well as on the wireless access links between the MC and the UEs. The emulator integrates OpenAirInterface's (OAI) implementation of the 5G Core Network (5GCN) and RAN, combined with RF Simulator [31] for exchanging time-domain samples within the RAN.

## 6.1    System deployment

The system designed for MC evaluation is illustrated in Figure 22. It consists of three main components: the Network namespace, the Mobile Cell namespace, and three UEs operating within their own namespaces. Different namespaces enable deploying the entire system on a single computer for validation purposes.

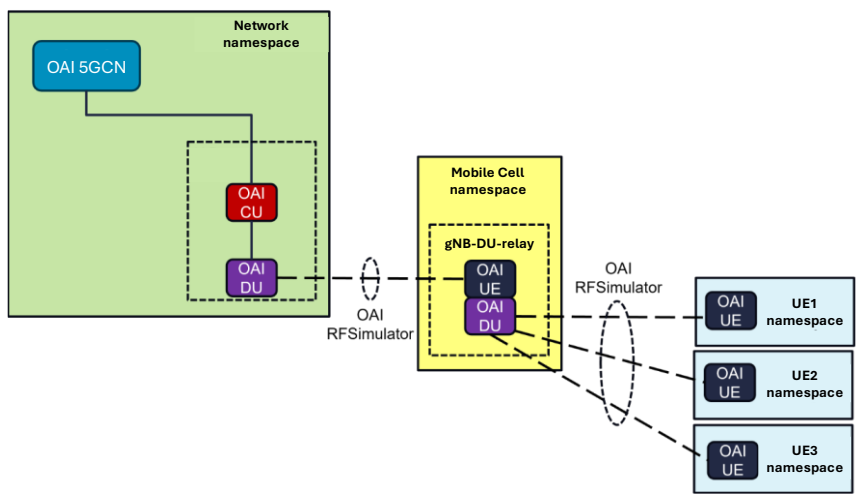

Figure 22. System deployed for Mobile Cell evaluation.

The Network namespace represents the MC's *Home Network* (HN) RAN entities and core functions. While the HN always plays the role of *Serving Network* (SN) for subscribed UEs that use the MC for access (not needing any roaming agreement), in the deployed set-up it also provides 5G backhaul connectivity, acting as *Overlay Network* (ON). Thus, it is possible to share the HN core functions (AMF and UPF) by the SN and ON logical network. In addition, we assume that the HN gNBs are based on the CU/DU split architecture; thus, a single gNB-CU may not only control stationary gNB-DUs but also MC-DUs. These possibilities were exploited, implying that the used OAI components (CU, DU and UE) may represent different SN and ON entities and play different roles, as described next.

The Network namespace hosts two OAI components (one CU and one DU), along with OAI's 5GCN. The 5GCN deployed in Docker containers was chosen for its simplicity and ability to meet functional requirements.

The OAI DU represents one ON gNB-DU, while the OAI CU plays two roles: it acts as an ON gNB-CU that controls the ON gNB-DU over an internal F1 interface, and as an MC-CU that controls the remote MC-DU over an independent F1 interface established across the ON.

The OAI DU (ON gNB-DU) includes an RF Simulator server to emulate the backhaul wireless channel. The RF Simulator is an OAI-based software framework that eliminates the need for physical Radio Frequency (RF) hardware, while providing a flexible, realistic environment for complex testing scenarios. Acting as a replacement for a real RF device, instead of transmitting signals over the air, it directly forwards samples between network interfaces.

The Mobile Cell namespace includes two OAI components (one UE and one DU). The OAI UE represents an MC-MT that performs regular UE procedures and connects to the ON gNB-DU through RF Simulator's channel emulation. The OAI DU represents the MC-DU, which establishes an F1 tunnel to the MC-CU using the MC-MT *PDU session* set up by the MC-MT through the ON. In order to enable the connection of multiple UEs, the MC-DU also includes an RF Simulator server. Modifications to RF Simulator enabled the independent control of channels for each UE and independent emulation of network conditions.

Each UE operates in its own namespace due to naming constraints in the OAI implementation of UEs. Connecting the Mobile Cell namespace to each UE namespace allows UEs to communicate through the MC. This architecture ensures system flexibility and scalability, accommodating multiple UEs.

All interfaces in Figure 18 (CP) and Figure 19 (UP) for the *mobile gNB-DU relay* system are implemented as virtual interfaces between the respective OAI components, including UEs and core functions, according to their roles. A virtual F1 interface was also implemented for the communication between the ON gNB-CU and gNB-DU.

The entire system was deployed on a single OMEN 15-en0017np laptop running Ubuntu 22.04. The OAI branch *2024.w09* was employed for the emulation environment. Telnet was used to manage channel conditions, with iPerf3 TCP sessions employed to assess network performance under varying conditions.

The channel emulation relies on dynamically adjusting RF Simulator's path loss variable in real time. For that purpose, the path loss for each link from the ON-gNB to the MC, as well as from the MC to the UEs, was calculated for each MC position, using the Urban Micro Cell (UMi) 3GPP channel model [32]. Transmission power was set to *48 dBm*, while noise was set to *−101 dBm*.

## 6.2    System validation – context and strategy

Laboratory tests and functional validation of the Mobile Cell prototype were initially carried out on the emulation platform and included preliminary evaluation of system performance. Throughput was chosen as a performance metric, since it allows assessing the impact of MAC and PHY layer mechanisms on system behaviour, as a first step for performance optimization that must meet specific requirements and criteria.

Throughput depends on how users are scheduled and frequency and time resources, in the form of *Physical Resource Blocks* (PRBs), are allocated to competing flows, and on the *Modulation and Coding Scheme* (MCS) used for transmission of *transport blocks*, which may have to be segmented into shorter equal-sized *code blocks*,

### Modulation and Coding Scheme

The performance of an MCS may be characterized by the *Block Error Ratio* (BLER) of *code blocks*, that is, the probability of having to retransmit erroneous *code blocks*. Thus, throughput decreases as BLER increases. The *transport* BLER is derived from the *code* BLER and both depend on channel conditions that vary over time.

*Channel quality* is mainly affected by *path los*s that varies with distance and contributes to reducing the signal power at the receiver, as well as by *interference* and *fading*. *Channel quality* is usually characterized by the *Reference Signal Received Power* (RSRP) and the *Signal to Interference and Noise Ratio* (SINR). RSRP represents the average received power of reference signals and is an indicator of signal strength; however, it does not account for interference or noise. SINR complements RSRP by quantifying signal quality as the ratio between the desired signal power and the combined interference and noise.

The following analysis applies to downlink transmissions; thus, the MC acts as *sender* and the UEs play the *receive*r role on the respective channels.

To reduce the need to retransmit erroneous *code blocks* and keep throughput as high as possible, the receivers provide the sender with a *Channel Quality Indicator* (CQI) for MCS selection.

The CQI is an integer with values from *0* (out of range) to *15* (highest quality). Each CQI value is associated with a specific MCS, characterized by a *modulation order* and a *code rate*, as specified by 3GPP in four CQI/MCS mapping tables [33] . The higher the CQI value, the higher the *MCS index*.

The sender uses the received CQI value as an index to the appropriate table to obtain (select) the MCS for downlink transmission.

A receiver computes the CQI for the measured SINR and reports the highest CQI value such that a target *transport* BLER is achieved for the corresponding MCS and *transport block size* (TBS). As defined in [33], the target *transport* BLER is 0.1 or 0.00001, depending on the CQI/MCS table used, while TBS is a function of the MCS and the number of *resource blocks* allocated for downlink transmission to the receiver.

Since SINR values are continuous, a CQI value represents a range of SINR values; the low threshold is the minimum SINR that still meets the target BLER for the same MCS and TBS. The high threshold is implicitly the low threshold for the next higher CQI value (and associated MCS). If a CQI/SINR look-up table were built mapping a CQI value to the respective SINR low threshold [34], the receiver would select the CQI value with the highest SINR threshold lower than the measured SINR.

Crossing an SINR threshold means that the current CQI value should be increased or decreased (as the case may be), thus mapping the new CQI value to a higher or lower *MCS index*, respectively.

However, in practice, adjacent SINR ranges partially overlap, in the sense that SINR variations in the vicinity of a threshold (due to random factors and MC or UE mobility) or imprecise computation of the SINR thresholds and/or of SINR measurement may create ambiguity in what concerns the right CQI value to report.

Whatever the reason, CQI (and *MCS index*) overestimation may lead to unacceptable BLER and a number of *code block* retransmissions, which would require reducing the *MCS index*. Conversely, CQI underestimation may guarantee stable operation (negligible or null BLER), but less efficient use of resources, that is, lower throughput than it would be possible if the next higher *MCS index* had been selected. Thus, failing to catch SINR variations in the vicinity of a threshold and reflect them on the CQI value may degrade throughput or miss an opportunity to increase throughput, respectively.

We now look to these problems from the sender perspective.

The sender selects an *MCS index* based on the CQI value reported by the receiver, implicitly assuming that the SINR value is within the receiver SINR budget that is compliant with the target BLER. However, the sender is not aware of the real SINR value nor of short-term variation that may increase BLER to an unacceptable value.

Nevertheless, the sender is informed (by means of negative acknowledgments) of *code blocks* that need to be retransmitted. Thus, it may calculate the average BLER over a transmission window and decide to reduce the *MCS index* if the measured BLER exceeds a critical value. Conversely, increasing the *MCS index* is, by default, triggered by an increase in the reported CQI value; in addition, the sender may attempt increasing the *MCS index*, as a means of channel probing, to assess whether it would result in a stable operation with null or negligible BLER.

The receiver may also help the sender to deal with these cases. Since the receiver has to detect erroneous *code blocks* to request their retransmission, it may calculate an average BLER and use this information together with the SINR value to provide a better estimation of the channel quality. If appropriate, it may reduce the CQI value so that the *MCS index* is reduced by the sender which should keep its own BLER calculation as backup. Similarly, when the SINR value is close to the threshold associated with the next higher CQI value and BLER is null, it may increase the CQI as a receiver-driven channel probing scheme, which is preferable to a blind probing triggered by the sender.

### Downlink scheduling

In 5G networks, scheduling and allocation of time and frequency resources for both uplink and downlink transmissions are performed by gNBs every *Transmission Time Interval* (TTI), which, in general, corresponds to the time-slot duration. For flexibility, different slot durations, which depend on the 5G numerology, are supported; the highest slot duration of *1 ms* applies to *15 kHz* subcarrier spacing. For low-latency services, scheduling may be performed on mini-slots and, thus, shorter TTIs.

When designing a scheduling algorithm, two conflicting objectives (*efficiency* and *fairness)* must be analysed, and establishing a compromise may be necessary. In this context, *Proportional Fair* (PF) scheduling is a reference when attempting the desired trade-off between these objectives.

To characterize its main attributes, it is useful to compare PF with other schedulers, as described in [35–37]. As references for efficiency and fairness, we selected *Maximum Throughput* (MT) and *Blind Equal Throughput* (BET) schedulers, respectively.

These schedulers use a priority function according to a specific metric to decide the flow(s) to be selected on each TTI. In the following, we use a unified notation that is equivalent to slightly different notations adopted in other papers.

The *Maximum Throughput* (MT) scheduler aims at maximizing system throughput. The MT algorithm is *channel-aware*, using the reported CQI to rank the UEs according to the respective *MCS index*

The priority metric applied to $UE_i$ is the instantaneous achievable data rate $d_i(t)$ on the respective downlink channel, which is derived from the selected MCS; the UE with the highest $d_i(t)$ in the current scheduling interval is selected for transmission.

Thus, UEs with comparatively low CQI have a lower probability of being scheduled and may even starve. This is unfair and, in general, not acceptable in 5G networks.

The *Blind Equal Throughput* (BET) scheduler attempts to provide *throughput fairness* among UEs, regardless of the respective channel quality, thus using a *channel-unaware* strategy. Fairness must be understood in the *max-min* sense: it is not possible to increase the throughput of any user without decreasing the throughput of another user with lower or equal throughput

The priority metric is the inverse of $R_i(t)$, which is the past average throughput of $UE_i$ estimated every TTI over a moving window that acts as an exponential low-pass filter.

In scheduling interval $t$, the UE with the lowest $R_i(t - 1)$ is selected for transmission. Then, $R_i(t)$ is updated for use in interval $(t + 1)$ as a weighted sum of the past average throughput $R_i(t - 1)$ and the achieved throughput $r_i(t)$ in the current TTI, with weights $(1 - \alpha)$ and $\alpha$, respectively, and $0 \leq \alpha \leq 1$ ($r_i(t) = 0$ for non-scheduled UEs):

$$R_i(t) = (1 - \alpha) \times R_i(t - 1) + \alpha \times r_i(t) \qquad (1)$$

On the one hand, giving higher scheduling priority to the UE with the lowest $R_i(t - 1)$, has the intention of increasing its average throughput due to the contribution of $r_i(t)$, in an attempt to level up the throughputs of all UEs. On the other hand, UEs with lower values of $r_i(t)$, due to poorer channel quality, will be scheduled more often, requiring more resources to achieve the same throughput as UEs that experience better channel quality. Thus, throughput fairness comes at the price of reducing system efficiency; the higher is the disparity in channel quality the lower is the efficiency.

Unlike *max-min fair* schedulers that aim at throughput fairness, the *Proportional Fair* (PF) scheduler tries to achieve *time fairness*. In the PF basic mode, the idea is to allocate, on average, the same amount of time-frequency resources on downlink channels to active UEs. However, due to differences in channel quality, the expected data rates $d_i(t)$ that can be achieved are likely different; thus, the instantaneous throughputs $r_i(t)$ achieved by UEs when scheduled and the estimated average throughputs $R_i(t)$ are also different

The objective is that the average throughputs are (ideally) scaled down by the same factor in relation to the instantaneous achievable data rate $d_i(t)$, which must be assessed for all UEs on each scheduling interval, since PF schedulers are *channel-aware*.

The idea of *proportional fairness* was introduced in [38] and PF schedulers have been used in CDMA and OFDMA systems, namely in 3G (UMTS) [39] and in 4G (LTE) [35,36] networks, and are candidates to support different applications in 5G networks [37]. Enhanced PF schedulers have been proposed in the literature, but are not discussed here, since they were not considered for the current phase of the work.

The PF priority metric combines MT and BET metrics, and is defined as the ratio of $d_i(t)$ to $R_i(t - 1)$ referred to scheduling interval $t$. Like other schedulers, the estimated average throughput $R_i(t)$ is updated using the same moving window as BET

On each TTI, the UE with the highest ratio is selected for transmission. After being updated, the average throughput $R_i(t)$ increases due to the term $\alpha \times r_i(t)$, thus reducing this ratio to promote the convergence to equal ratios (ideally, $R_i(t)$ should converge to the average of $r_i(t)$).

Usually, the moving average formula is presented on a different way, defining $\alpha$ as the inverse of the window size $w$:

$$R_i(t) = (1 - 1/w) \times R_i(t - 1) + (1/w) \times r_i(t) \qquad (2)$$

In this sense, the window duration $t_w = w \times TTI$ may be used to control the access delay of UEs, which justifies that $\alpha$ is often expressed as the inverse of $t_w$, implicitly assuming $TTI = 1\ ms$, but this may not be the case in 5G; thus, $\alpha$ should be expressed as the ratio of TTI to $t_w$. In addition, $t_w$ controls the rxate of decrease of $R_i(t)$ (first term of the sum) of all UEs, and the decrease, on the next TTI, of the access probability of UE(s) scheduled in the current TTI, since a non-null $r_i(t)$ (second term of the sum) increases the respective $R_i(t)$ that is used in interval $(t+1)$.

For very high values of $w$ ($\alpha \rightarrow 0$), the PF scheduler behaves as an MT scheduler, since the weight of $r_i(t)$ becomes negligible (second term of the sum) and the UE with the best channel quality is scheduled with higher priority.

When $w \rightarrow 1$ ($\alpha \rightarrow 1$), the gain of using CQI information is lost, since the past average throughput $R_i(t)$ exponentially converges to $0$, even when updated with the addition of $\alpha \times r_i(t)$ after $UE_i$ is scheduled (for example, if $\alpha = 0.99$, the first term of the sum decreases by two orders of magnitude on each TTI). Thus, the scheduler quickly forgets past information (in the limit, with $\alpha = 0$, $R_i(t) = r_i(t)$, which is $0$, except when $UE_i$ is scheduled). In practice, the scheduler is not capable of dealing with sudden changes of channel quality, by not reflecting and keeping in $R_i(t)$ the effect of changes in the achievable data rate $d_i(t)$ and, thus, on the instantaneous achieved throughput $r_i(t)$.

### Validation goals and traffic scenarios

In laboratory tests, we used OAI procedures to select the *MCS index* and the OAI PF scheduler [40]. In this sense, the throughput study was not targeted at optimizing performance (by fine-tuning critical parameters) but as part of functional validation of the MC prototype, namely: 1) confirming the correct establishment of IP connectivity over an F1 interface and 2) verifying whether the execution of the OAI algorithms produced consistent variation of throughput per UE flow with SINR, by placing the MC on a set of positions along a path, thus changing the distance to each UE with three different patterns; UE throughput was assessed in two traffic scenarios.

In the first scenario, only one UE was active at a time, receiving a TCP flow transferred through the MC; simulations were run for each UE separately. This allowed, for each MC position, deriving the average $UE_i$ throughput from the instantaneous achieved throughput $r_i(t)$ on all scheduling intervals, which is limited by the achievable data rate $d_i(t)$ expected with the *MCS index* selected on each TTI. The throughput results in this scenario are used as reference, since all resources (subcarriers and time slots) are allocated by the OAI PF scheduler to a single UE.

In the second scenario, all UEs were active receiving independent TCP flows on the respective channels. Resources are allocated to competing UEs using the PF priority metric; thus, the average throughput per UE was obtained from the $r_i(t)$ values on the intervals in which the respective flows were scheduled for transmission. Comparing these results with those of the first scenario, it is possible assessing the behaviour of the scheduling algorithm, as well as validating the correct execution of the critical MAC and PHY procedures supported by the OAI implementation.

The validation scenario, depicted in Figure 23, employed a *3.6 GHz* carrier frequency and a *20 MHz* channel bandwidth.

In this set-up, the ON-gNB and UEs remained stationary, while the MC was sequentially placed at 21 different positions, from (*1000, 25*) *m* to (*1990, 1225*) *m*. The antenna heights were set at *15 m* for both the ON-gNB and the MC, and at *2 m* for the UEs. The MC is assumed to be a van equipped with a deployable antenna mast, typically used when the vehicle is stationary.

To assess system performance, we considered three UEs connected to the network through the Mobile Cell. Traffic consisted of TCP flows in the downlink direction, originated by a container within the 5GCN and directed to each UE. TCP traffic generated by iPerf3 enables sustained data transfer over the duration of each test, allowing for accurate measurement of network throughput by utilizing the available bandwidth under controlled conditions.

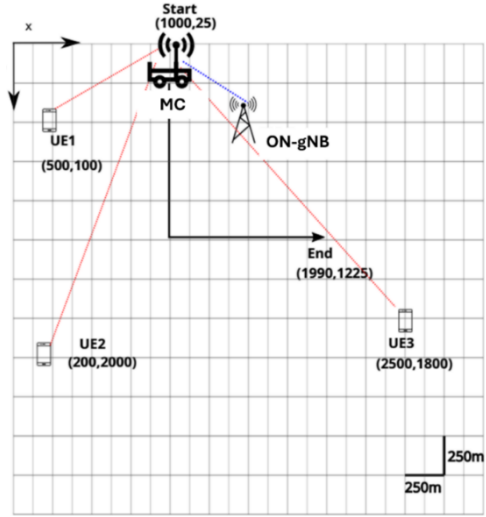

Figure 23. Validation scenario.

## 6.3    Performance evaluation

Throughput on the downlink channel depends, for each UE, on the SINR and, thus, on the received signal power, which decreases as the distance to the MC increases. For this reason, we first present in Figure 24 the *Reference Signal Received Power* (RSRP) measurements at each UE for all 21 MC positions.

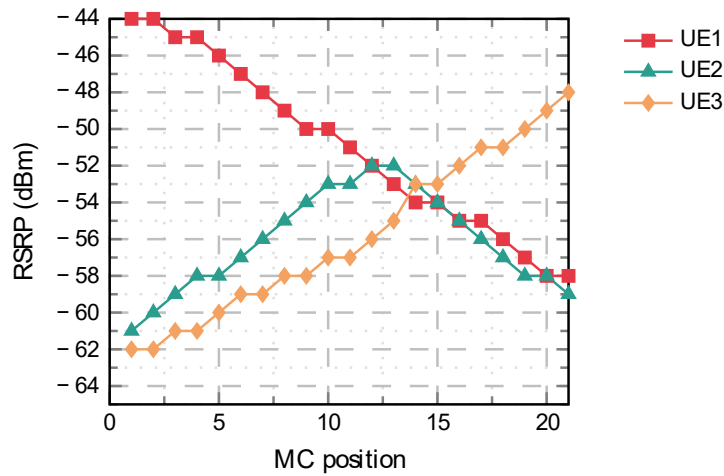

Figure 24. RSRP measured on all UEs for each MC position.

The figure illustrates three different patterns: RSRP gradually decreases for UE1, first increases and then decreases for UE2, and increases all the way for UE3, since the distances vary the opposite way as the MC moves along the predefined path.

Thus, a steady increase of RSRP should provoke a step-by-step increase of throughput, whenever possible to select a higher *MCS index*; conversely, a step-by-step decrease of throughput should occur whenever selecting a lower *MCS index*, due to an RSRP decrease below a critical threshold.

In the simulations, MCS selection and PF scheduling were automatically defined by the default OAI mechanisms and, thus, not optimized.

In Scenario 1, in which there is no competition for resources and, thus, the PF scheduler is not invoked, this trend is confirmed, as illustrated in Figure 25. However, two comments are still required.

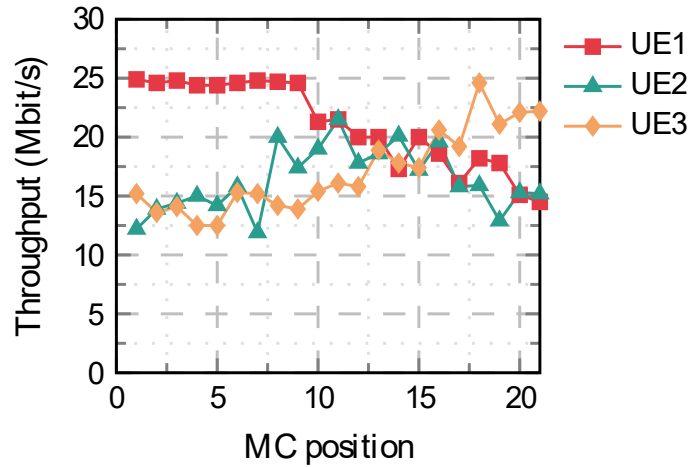

Figure 25. Scenario 1 – comparison of UE throughput.

Firstly, the throughput of UE1 only starts to consistently decrease around position 10. This means that, until then, the RSRP value was high enough to keep selecting the highest possible *MCS index*.

Secondly, all curves show throughput fluctuations, which may be explained by a non-optimum selection of the *MCS index*, due to inaccurate CQI estimation on critical RSRP values, as discussed. Thus, throughput degradation may occur when selecting a too high *MCS index* that leads to MAC retransmissions and even losses, in which case TCP retransmissions are also necessary; conversely, opportunities to increase throughput may be missed when selecting a lower *MCS index* than would be possible.

In Scenario 2, resources of the downlink shared channel are allocated to competing UEs by a PF scheduler. Throughput results are shown in Figure 26. The following conclusions may be derived.

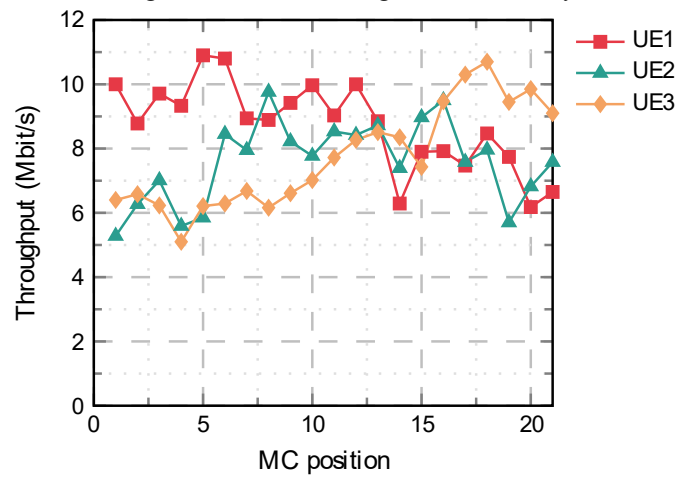

Figure 26. Comparison of the throughput for UE1, UE2, and UE3.

As intended, UE throughput values are scaled down. Throughput fluctuations are also observed and may have been aggravated due to the PF scheduler failing to achieve more accurately the fairness objective. Such fluctuations are more evident than in the first scenario sine the amplitudes of UE throughput values are also scaled down, but the same trend in what concerns the variation of throughput with RSRP is perceived.

Moreover, *Modulation and Coding Scheme* selection and *Proportional Fair* scheduling were automatically defined by the default OAI mechanisms and, thus, not optimized.

The results complete the functional validation of the MC implemented as a *mobile gNB-DU relay* based on the overlay model and provide a preliminary assessment of the system performance in two traffic scenarios.

It is worth noting that the system performance in the emulated scenarios using the RF Simulator may be worse than might be achieved in real-world deployments. This presumption is mainly justified by the processing constraints inherent to the emulation environment. Unlike physical deployments, where dedicated RF hardware enables real-time sampling and processing, the RF Simulator operates in a CPU-bound environment. In this set-up, sample exchanges and channel emulation processes rely on available computational resources. Limited CPU resources available over time reduce processing capabilities, resulting in throughput measurements that are constrained by simulation efficiency rather than actual network capacity. Consequently, the observed performance in emulation reflects a worst-case estimation, influenced by computational overheads, rather than the full potential of the network in a real-world, hardware-accelerated environment.

Since performance evaluation was carried out with a real implementation of OAI, the network set-up configurations developed in the emulated environment are fully portable to real-world deployments. The use of OAI ensures that the protocol stack used in the emulated environment is consistent with those in real deployments. This highlights the relevance of the experimental results presented herein. While the RF simulator introduces processing constraints that limit throughput in the emulated environment, the functionality and architecture of the 5G system remain unchanged. Consequently, the tested set-up can be directly applied to physical RF hardware, either based on *Software-Defined Radio* (SDR) or *Radio Units* (RUs), where real-time sampling and optimized processing enable the network to achieve its full performance.

The use of the real OAI implementation in the set-up presented in this paper bridges the gap between design and deployment of a 5G MC.

# 7    Conclusions and future work

This paper introduced the concept of a 5G Mobile Cell (MC) aimed at offering flexible, on-demand wireless connectivity to extend coverage or reinforce capacity of an operational public or private 5G network or to provide services where a permanent 5G access infrastructure does not exist, is insufficient or damaged.

As an alternative to a *Mobile Base Station Relay* (MBSR) specified by 3GPP based on the *Integrated Access and Backhaul* (IAB) architecture, we favoured two MC solutions that rely on a 5G overlay network for backhauling between the mobile system and the MC home network composed of a stationary RAN and core functions, Using a modular design, the MC may be deployed as a *mobile gNB*, compliant with the *Wireless Access Backhaul* (WAB) architecture, being specified by 3GPP, or as a *mobile gNB-DU relay*, which is based on the CU/DU split gNB architecture, like the IAB MBSR.

Based on the respective protocol stacks, we carried out a comparative analysis of the three solutions and discussed trade-offs between complexity, backhaul integration, and performance requirements. This comparison, complemented with experimental and theoretical results of on-going work, may help network operators and service providers in choosing the MC architecture that best meets deployment, service and performance requirements of specific target application scenarios envisaged for the MC.

The OAI real implementation was used for emulation-based evaluation of the MC as a *mobile gNB-DU relay*. Simulation results confirmed the impact of the MC position on performance; UE throughput depends on channel quality, and both degrade as the distance to the MC increases. In this context, the channel-aware PF scheduler proved to be a good compromise between efficiency and fairness. We also concluded that, to improve performance, optimization of MAC and PHY algorithms is still necessary,

Current work is focused on integrating the MC prototype, which will be deployed and tested on a real 5G pilot for validation of the simulation results and proof of concept. This will allow testing the MC in more realistic propagation conditions and dynamic constrained environments, with varying UE mobility patterns and traffic demand. The emulation platform will continue to be used for more advanced simulations.

On the one hand, a better characterization of the MC architectures based on the overlay model requires analysing and evaluating the impact on throughput and delay of the backhaul network, as well as supporting the F1 interface across this network.

On the other hand, optimization of system performance will address two issues discussed in Section 6.2: 1) enhancement of the MCS selection procedures, both on the sender and receiver sides, as outlined there, and

2) improvement of the PF scheduler, e.g. the convergence of the moving window estimation of the average throughput, taking advantage of the way PRBs are allocated to scheduled UEs in 5G networks.

For application scenarios that require installing the MC in a predefined (quasi) static location, we plan to employ state-of-the-art algorithms and optimization techniques for dynamically adjusting the MC position and configuration based on real-time network metrics. By integrating an adaptive placement and network management approach, the MC holds the potential to provide reliable connectivity in diverse operational scenarios. Future research work may extend this analysis to dynamic real-world environments, accounting for varying UE mobility patterns and traffic demand

## Acknowledgments

This work is co-financed by Component 5-Capitalization and Business Innovation, integrated in the Resilience Dimension of the Recovery and Resilience Plan within the scope of the Recovery and Resilience Mechanism (MRR) of the European Union (EU), framed in the Next Generation EU, for the period 2021-2026, within project NEXUS, with reference 53.

## References

1. Jaffry, S.; Hussain, R.; Gui, X.; Hasan, S.F. A Comprehensive Survey on Moving Networks. *IEEE Communications Surveys & Tutorials* **2021**, *23*, 110–136, doi:10.1109/COMST.2020.3029005.
2. Pozza, M.; Rao, A.; Flinck, H.; Tarkoma, S. Network-In-a-Box: A Survey About On-Demand Flexible Networks. *IEEE Communications Surveys & Tutorials* **2018**, *20*, 2407–2428, doi:10.1109/COMST.2018.2807125.
3. Almeida, E.N.; Coelho, A.; Ruela, J.; Campos, R.; Ricardo, M. Joint Traffic-Aware UAV Placement and Predictive Routing for Aerial Networks. *Ad Hoc Networks* **2021**, *118*, 102525, doi:10.1016/j.adhoc.2021.102525.
4. Li, J.; Nagalapur, K.K.; Stare, E.; Dwivedi, S.; Ashraf, S.A.; Eriksson, P.-E.; Engström, U.; Lee, W.-H.; Lohmar, T. 5G New Radio for Public Safety Mission Critical Communications. *IEEE Communications Standards Magazine* **2022**, *6*, 48–55, doi:10.1109/MCOMSTD.0002.2100036.
5. NEXUS – Innovation Agenda for Digital and Green Transition. Available online: https://nexus-uslab.pt/.
6. Szymanowska, B.B.; Kozłowski, A.; Dąbrowski, J.; Klimek, H. Seaport Innovation Trends: Global Insights. *Mar Policy* **2023**, *152*, 105585, doi:https://doi.org/10.1016/j.marpol.2023.105585.
7. Charpentier, V.; Slamnik-Kriještorac, N.; Landi, G.; Caenepeel, M.; Vasseur, O.; Marquez-Barja, J.M. Paving the Way towards Safer and More Efficient Maritime Industry with 5G and Beyond Edge Computing Systems. *Computer Networks* **2024**, *250*, 110499, doi:https://doi.org/10.1016/j.comnet.2024.110499.
8. Du, R.; Mahmood, A.; Auer, G. Realizing 5G Smart-Port Use Cases with a Digital Twin. *Ericsson Technology Review* **2022**, *2022*, 2–11, doi:10.23919/ETR.2022.9985778.
9. Coelho, A.; Ruela, J.; Queirós, G.; Trancoso, R.; Correia, P.F.; Ribeiro, H.; Fontes, H.; Campos, R.; Ricardo, M. On-demand 5G Private Networks using a Mobile Cell. Accepted at the 2024 NEXUS International Conference (DGTMP 2024); 2024; pp. 1–6. Available online: https://arxiv.org/abs/2411.06597.
10. 3GPP System Architecture for the 5G System (Release 18) 2024, *TS 23.501*.
11. 3GPP NR; NR and NG-RAN Overall Description; Stage 2 (Release 18) 2024, *TS 38.300*.
12. 3GPP Study on New Radio Access Technology: Radio Access Architecture and Interfaces (Release 14) 2017, *TS 38.801*.
13. Alliance, O. O-RAN Architecture Description. *O-RAN.WG1.OAD-R003-v12.00, Technical Specification* **2024**.

14. Polese, M.; Bonati, L.; D'Oro, S.; Basagni, S.; Melodia, T. Understanding O-RAN: Architecture, Interfaces, Algorithms, Security, and Research Challenges. *IEEE Communications Surveys & Tutorials* **2023**, *25*, 1376–1411, doi:10.1109/COMST.2023.3239220.

15. Ericsson AB; others *Common Public Radio Interface: ECPRI Interface Specification*; 2019;

16. 3GPP NG-RAN; Architecture Description (Release 18) 2024, *TS 38.401*.

17. 3GPP NG-RAN; F1 General Aspects and Principles (Release 18) 2024, *TS 38.470*.

18. 3GPP NG-RAN; F1 Application Protocol (F1AP) (Release 18) 2024, *TS 38.473*.

19. Stewart, R.R.; Tüxen, M.; karen Nielsen Stream Control Transmission Protocol 2022.

20. 3GPP General Packet Radio System (GPRS) Tunnelling Protocol User Plane (GTPv1-U) (Release 18) 2023, *TS 29.281*.

21. Zhang, Y.; Kishk, M.A.; Alouini, M.-S. A Survey on Integrated Access and Backhaul Networks. *Frontiers in Communications and Networks* **2021**, *2*, doi:10.3389/frcmn.2021.647284.

22. 3GPP NR; Backhaul Adaptation Protocol (BAP) Specification (Release 18) 2024, *TS 38.340*.

23. 3GPP Service Requirements for the 5G System; Stage 1 (Release 19) 2024, *TS 22.261*.

24. 3GPP Architecture Enhancements for 5G System (5GS) to Support Vehicle-to-Everything (V2X) Services (Release 18) 2024, *TS 23.287*.

25. 3GPP Procedures for the 5G System (5GS); Stage 2 (Release 18) 2024, *TS 23.502*.

26. 3GPP Policy and Charging Control Framework for the 5G System (5GS); Stage 2 (Release 18) 2024, *TS 23.503*.

27. NTT DOCOMO, I.N.C.; AT&T *Revised SID: Study on Additional Topological Enhancements for NR*; 2024;

28. 3GPP Study on Additional Topological Enhancements for NR (Release 19) 2024, *TR 39.799*.

29. 3GPP NR; Radio Resource Control (RRC) Protocol Specification (Release 18) 2024, *TS 38.331*.

30. Cojocaru, I.; Coelho, A.; Ricardo, M. Mobile Node Emulator for 5G Integrated Access and Backhaul Networks. In Proceedings of the 2024 20th International Conference on Wireless and Mobile Computing, Networking and Communications (WiMob); 2024; pp. 599–602.

31. OpenAirInterface (OAI) Project OpenAirInterface 5G RF Simulator 2024.

32. ETSI Study on Channel Model for Frequencies from 0.5 to 100 GHz (3GPP TR 38.901 Version 18.0.0 Release 18) 2024.

33. 3GPP *TS 38.214 V18.5.0; 5G; NR; Physical Layer Procedures for Data (Release 18)*; 2025;

34. Lagen, S.; Wanuga, K.; Elkotby, H.; Goyal, S.; Patriciello, N.; Giupponi, L. New Radio Physical Layer Abstraction for System-Level Simulations of 5G Networks. In Proceedings of the ICC 2020 - 2020 IEEE International Conference on Communications (ICC); 2020; pp. 1–7.

35. Kela, P.; Puttonen, J.; Kolehmainen, N.; Ristaniemi, T.; Henttonen, T.; Moisio, M. Dynamic Packet Scheduling Performance in UTRA Long Term Evolution Downlink. In Proceedings of the 2008 3rd International Symposium on Wireless Pervasive Computing; 2008; pp. 308–313.

36. Capozzi, F.; Piro, G.; Grieco, L.A.; Boggia, G.; Camarda, P. Downlink Packet Scheduling in LTE Cellular Networks: Key Design Issues and a Survey. *IEEE Communications Surveys & Tutorials* **2013**, *15*, 678–700, doi:10.1109/SURV.2012.060912.00100.

37. Haque, Md.E.; Tariq, F.; Khandaker, M.R.A.; Wong, K.-K.; Zhang, Y. A Survey of Scheduling in 5G URLLC and Outlook for Emerging 6G Systems. *IEEE Access* **2023**, *11*, 34372–34396, doi:10.1109/ACCESS.2023.3264592.

38. Kelly, F. Charging and Rate Control for Elastic Traffic. *European Transactions on Telecommunications* **1997**, *8*, 33–37, doi:https://doi.org/10.1002/ett.4460080106.

39. Bu, T.; Li, L.; Ramjee, R. Generalized Proportional Fair Scheduling in Third Generation Wireless Data Networks. In Proceedings of the Proceedings IEEE INFOCOM 2006. 25TH IEEE International Conference on Computer Communications; 2006; pp. 1–12.

40. Ursu, R.-M.; Papa, A.; Kellerer, W. Experimental Evaluation of Downlink Scheduling Algorithms Using OpenAirInterface. In Proceedings of the 2022 IEEE Wireless Communications and Networking Conference (WCNC); 2022; pp. 84–89.